\documentclass[12pt,preprint]{aastex}
\accepted{}
\journalid{}{}
\articleid{}{}

\lefthead{Mirabal {\it et al.}}
\righthead{GRB 021004: A possible Shell Nebula around a GRB Progenitor}

\received{2002 March 4}
\begin{document}

\def\etal{{\it et al.}}
\def\eg{{e.g.,}}
\def\ie{{i.e.,}}
\def\vs{{\it vs.}}
\def\etc{{\it etc.}}
\def\kms{km~s$^{-1}$}
\def\Msol{M$_\odot$}
\def\lsim{\mathrel{\lower .85ex\hbox{\rlap{$\sim$}\raise
.95ex\hbox{$<$} }}}
\def\gsim{\mathrel{\lower .80ex\hbox{\rlap{$\sim$}\raise
.90ex\hbox{$>$} }}}
\newbox\grsign \setbox\grsign=\hbox{$>$}
\newdimen\grdimen \grdimen=\ht\grsign
\newbox\laxbox \newbox\gaxbox
\setbox\gaxbox=\hbox{\raise.5ex\hbox{$>$}\llap
     {\lower.5ex\hbox{$\sim$}}}\ht1=\grdimen\dp1=0pt
\setbox\laxbox=\hbox{\raise.5ex\hbox{$<$}\llap
     {\lower.5ex\hbox{$\sim$}}}\ht2=\grdimen\dp2=0pt
\def\gax{\mathrel{\copy\gaxbox}}
\def\lax{\mathrel{\copy\laxbox}}

\def\pz{\phantom{0}}
\def\lsim{\mathrel{\lower .85ex\hbox{\rlap{$\sim$}\raise
.95ex\hbox{$<$} }}}
\def\gsim{\mathrel{\lower .80ex\hbox{\rlap{$\sim$}\raise
.90ex\hbox{$>$} }}}
\def\bv{($sB-V$)}
\def\vr{($V-R$)}
\def\br{($B-R$)}
\def\ub{($U-B$)}
\def\vi{($V-I$)}
\def\ri{($R-I$)}
\def\source{3EG~J1835+5918}
\def\ro{{\it ROSAT\/}}
\def\asca{{\it ASCA\/}}
\newcommand{\ia}{\'\i}

\title{GRB 021004: A Possible Shell Nebula around a Wolf-Rayet 
Star Gamma-Ray Burst Progenitor\altaffilmark{1}}

\author{N. Mirabal\altaffilmark{2}, J. P. Halpern\altaffilmark{2}, 
Ryan Chornock\altaffilmark{3}, Alexei V. Filippenko\altaffilmark{3},
D. M. Terndrup\altaffilmark{4}, E. Armstrong\altaffilmark{2}, J. Kemp\altaffilmark{2,5}, J. R. Thorstensen\altaffilmark{6}, M. Tavarez\altaffilmark{7},
 \&
C. Espaillat\altaffilmark{2}}
\altaffiltext{1}{Based in part on data obtained at the W.M. Keck Observatory,
which is operated as a scientific partnership among the California 
Institute of Technology, the University of California, and NASA, and
was made possible with the generous financial support of the W.M. Keck 
Foundation}
\altaffiltext{2}{Astronomy Department, Columbia University, 550 West 120th 
Street, New York, NY 10027}
\altaffiltext{3}{Department of Astronomy, 601 Campbell Hall, University
        of California, Berkeley, CA 94720-3411}
\altaffiltext{4}{Department of Astronomy, Ohio State University, Columbus, OH 43210} 
\altaffiltext{5}{Joint Astronomy Centre, University Park, 660 North A'ohoku Place, Hilo, HI 96720}
\altaffiltext{6}{Department of Physics and Astronomy, Dartmouth College, 6127 Wilder Laboratory, Hanover, NH 03755-3528}
\altaffiltext{7}{Department of Astronomy, University of Michigan, Ann Arbor, MI 48109}

\begin{abstract}
\rightskip 0pt \pretolerance=100 \noindent

The rapid localization of GRB 021004 by the HETE-2 satellite allowed nearly continuous monitoring of its early
optical afterglow decay, as well as high-quality optical spectra that determined a redshift of $z_{3}$=2.328 for its host galaxy, an active
starburst galaxy with strong Lyman-$\alpha$ emission and several 
absorption lines. Spectral observations show  multiple
absorbers at $z_{3A}= 2.323$,
$z_{3B}= 2.317$, and $z_{3C}= 2.293$ blueshifted by $\sim$ 450, 
$\sim$ 990, and  $\sim$ 3,155 km~s$^{-1}$ respectively 
relative to the host galaxy Lyman-$\alpha$ 
emission. We argue that these correspond to a 
fragmented shell nebula that 
has been radiatively accelerated by the gamma-ray burst (GRB) afterglow
 at a distance $\gax$ 0.3 pc from a Wolf-Rayet star GRB progenitor. 
The chemical abundance ratios 
indicate that the nebula is overabundant
in carbon and silicon. 
The high level of carbon and silicon is consistent
with a swept-up  
shell nebula gradually enriched by a WCL progenitor wind over the 
lifetime of the nebula 
prior to the GRB onset. The detection of statistically
significant fluctuations and color changes about the 
jet-like optical decay 
further supports this interpretation since 
fluctuations must be present at some level 
due to irregularities in a clumpy stellar wind
medium or if the progenitor has undergone massive ejection prior to the
GRB onset. This evidence suggests that the 
mass-loss process in a Wolf-Rayet   
star might lead naturally 
to an iron-core collapse with sufficient angular momentum 
that could serve as a suitable GRB progenitor. 
Even though we cannot rule out definitely 
the alternatives of a dormant QSO, large-scale superwinds, or 
a several hundred 
year old supernova remnant responsible for the blueshifted absorbers, these
findings point to the possibility of a likely signature for a  
massive-star GRB progenitor.
\end{abstract}


\section{Introduction}

Considerable evidence exists 
connecting long-duration GRBs to 
star-forming regions and consequently to
a massive-star origin. For instance, 
optical spectroscopy of well-calibrated 
emission lines has been used to derive star-formation rates (SFRs) that place
GRB host galaxies slightly above the field galaxy population
at comparable redshifts, in terms of SFR (Djorgovski et al. 
2001). GRB locations within their host 
galaxies also seem to follow closely the galactic light distribution and  
are hard to reconcile with coalescing compact objects in a galactic halo
(Bloom, Kulkarni, \& Djorgovski 2002). Additional clues have come
from secondary peaks observed
in the late-time optical light curves of a few GRBs that
have been interpreted as supernova (SN) emission associated with the 
GRB formation
(\eg~Bloom et al. 2002; Garnavich et al. 2003).
Recently, spectra of the GRB 030329 afterglow have shown 
an emergence of broad
features characteristic of the peculiar type-Ic 
supernovae (Stanek et al. 2003; Chornock et al. 2003). 
Driven by the observational evidence and detailed calculations,  
two models have emerged as the leading massive-star GRB progenitors, 
namely, collapsars and supranovae. 
The collapsar model (Woosley 1993; 
MacFadyen \& Woosley 1999)  
corresponds to a black hole formed promptly in a 
massive-star core-collapse (typically a
Wolf-Rayet star) that fails to produce a successful  
outgoing shock (Type I), or in the less extreme case 
a ``delayed black hole'' results by  
fallback after a weak outgoing shock (Type II). In the supranova model, a GRB 
takes place once the centrifugal
support of a ``supramassive'' neutron star, formed months or years prior to
the event, weakens and the neutron 
star collapses to form a black hole (Vietri \& Stella 1998). 

Although an association with massive-star collapse was among
the first theories proposed to explain GRBs (Colgate 1974), 
a definite local signature of the GRB progenitor is still being sought.
The recent detection of 
blueshifted H, C IV, and Si IV  
absorbers in the spectrum of the GRB 021004 
afterglow (Chornock \& Filippenko 2002), coupled with the
irregularities observed in its optical light curve, has been interpreted 
as evidence
of a clumpy wind from a massive-star progenitor, such as a WC Wolf-Rayet
star (Mirabal et al. 2002a; Schaefer et al. 2003). 
In this paper,
we discuss what might constitute the first detection of a fragmented 
shell nebula around
a GRB progenitor. Our basic approach in this analysis is to begin with
simple models consistent with the
photometry and spectroscopy of the GRB 021004 afterglow. We
then consider the physical parameters for each model and 
introduce modifications that best fit the GRB 021004 data. 
The outline of the paper
is as follows: \S 2 describes the optical photometry and 
spectroscopy, while  \S 3 describes
the temporal decay, broadband modeling of the afterglow, 
absorption-line identification,  
and abundance analysis. In \S 4 
and \S 5, we detail
the evolution of a massive-star shell nebula and 
radiative acceleration models. An in-depth analysis 
of alternative explanations is given in \S 6. Finally,  
the implications of our results for GRB progenitors are 
presented in \S 7, and \S 8 summarizes our conclusions.

\section{Observations}

\subsection{Optical Photometry}

GRB 021004 is to date the fastest localized long-duration GRB detected
by the HETE-2 satellite (Shirasaki et al. 2002). The HETE-2 FREGATE, WXM, and
SXC instruments detected the event on 2002 Oct. 4.504 (UT dates
are used throughout this paper) with 
a duration of $\approx$100~seconds. 
The improved flight localization software 
in the WXM instrument produced a reliable 
position only 49 seconds after the beginning of the burst, that was later
refined by the ground analysis. Rapid follow-up 
detected a bright optical transient (OT) 
inside the 90$\%$ WXM confidence circle only 10 minutes after the initial
HETE-2 notice (Fox 2002). 

We began optical observations of the OT
14.7 hr after the burst by obtaining 
an equal number of well-sampled, high signal-to-noise ratio 
$B$, $V$, $R$, and $I$ images using the 1.3 m
and 2.4 m telescopes at the MDM Observatory
(Halpern et al. 2002). Nearly nightly  
observations were carried out in the $B$ and $R$ bands 
until 2002 Oct. 25 with additional late-time 
measurements on 2002 Nov. 25-27. We placed all the optical
observations on a common $BVRI$ system using the 
latest calibration of nearby field stars acquired by Henden (2002). 
The MDM photometric measurements 
including errors are listed in Table 1 and shown in Figure 1.
For clarity in Figure 1 we have omitted 
the early-time observations, \ie~$t \lax$ 14.7 hr after
the burst (refer to Fox et al. 2003 for details). 

\subsection{Optical Spectroscopy}

Optical spectra were obtained with the dual-beam
Low Resolution Imaging Spectrometer (LRIS; Oke et al. 1995) on the Keck-I
10 m telescope on 2002 Oct. 8.426-8.587 (Chornock \& Filippenko 2002).
The spectra were taken in five individual 1200 s exposures using 
a $1^{\prime\prime}$ wide slit.  The skies were variably cloudy, so the
first three exposures were of noticeably higher quality than the last two.
We used a 400 lines/mm grating blazed at 8500 \AA\ on the red side and 
a 400 lines/mm grism blazed at 3400 \AA\ on the blue side. 
The effective spectral resolution is $\sim$ 6 \AA~~on both 
the blue and red sides.
The data were trimmed, bias-subtracted, and flat-fielded using
standard procedures. Extraction of the spectra was performed using IRAF
\footnote{IRAF is distributed by the National Optical Astronomy Observatories,
    which are operated by the Association of Universities for Research
    in Astronomy, Inc., under cooperative agreement with the National
    Science Foundation.}.
The wavelength scale was established by fitting polynomials to Cd-Zn and 
Hg-Ne-Ar lamps.
Flux calibration was accomplished using our own IDL procedures 
(Matheson et al. 2000) and comparison exposures of the spectrophotometric
standard stars BD +28$^{\circ}$ 4211 and BD +17$^{\circ}$ 4708 on the 
blue and red sides, 
respectively (Stone 1977; Oke \& Gunn 1983).
We removed the atmospheric absorption bands through division by the
intrinsically smooth spectra of the same standard stars
(Matheson et al. 2000).  The two halves of the spectrum were averaged
in the 5650-5700 \AA\ overlap region.

\section{Analysis}

\subsection{Temporal Decay and Environment}

Early analysis of the OT revealed 
statistically significant fluctuations about a simple 
power-law decay (Bersier et al. 2003; Halpern et al. 2002). 
Although the general trend of 
the early optical decay can be fitted by a simple power-law fit, shown in 
Figure 2, significant deviations about the mean decay are present on time 
scales from minutes to hours. Figure 3 also shows a distinct
color change starting around 1.6 days after the burst in agreement with
the results reported by Bersier et al. (2003).
It has been postulated that 
deviations from a simple power-law behavior might be induced by
inhomogeneities in the circumburst medium (Wang \& Loeb 2000),
structure within a jet (Kumar \& Piran 2000),
and/or if the afterglow is ``refreshed'' by collisions among separate
shells (Rees \& M\'esz\'aros 1998). 
The possible causes of the deviations and color changes
in the GRB 021004 OT will be discussed at greater length in \S 7.

By day 9, the gradual decay of the
OT became clearly inconsistent with the early-time power-law
fit and turned steeper in its decay slope. 
In order to describe the steepening of the afterglow decay, we
fitted the data with a smooth function taking into account a constant 
host-galaxy contribution and a broken power-law behavior of the form
\begin{equation} 
 F(t)\ =\ {2\,F_b\,(t/t_b)^{\alpha_1} \over 
1+(t/t_b)^{(\alpha_1-\alpha_2)}}\ +\ F_0, 
\end{equation}
\noindent
where $\alpha_1$ and $\alpha_2$ represent the 
asymptotic early and late-time
slopes, $F_0$ is the constant galaxy contribution, 
and $F_b$ is the OT flux at the break time $t_b$ (Halpern et al. 2000).
The best fit to the data is found for
$\alpha_{1} = -0.72$, $\alpha_{2} = -2.9$, and 
$t_{b}$ = 9 days. In Figure 1, we draw the fit including the constant
contribution of the host galaxy which contaminates the OT 
at late times.  The   
host galaxy contribution was determined from 
deep $B$ and $R$ imaging
obtained on 2002 Nov. 25-27 under good seeing conditions.  
The images reveal a relatively blue
host galaxy, $(B-R)_{host} \approx$ 0.65 mag, with estimated
magnitudes  $R_{host}$ = 23.95 $\pm 0.08$ and $B_{host}$ = 24.60 $\pm 0.06$,
measured in an aperture that includes the total contribution of the 
host galaxy. The estimated host galaxy
color is bluer than the OT itself [$(B-R)_{OT} \approx$ 1.05 mag] 
and bluer than
nearby field galaxies. Figure 4 shows 
images of the GRB 021004 OT at early ($t \approx$ 19.8 hr) and 
late ($t \approx$ 52 days) times when the
host galaxy dominates.  A recently released (HST Program
9405, PI: Fruchter) high-resolution 
image of the OT obtained 
with the Advanced Camera for Surveys (ACS) on
{\it HST\/} with the F606W filter, shown in Figure 5, 
confirms the emergence of an underlying
host galaxy by 2002 Nov. 26. Unfortunately, it is difficult
to resolve the contribution from the OT cleanly (Levan et al. 2003).

The early-time optical photometry of the OT, 
in comparison with the X-ray flux obtained 
0.85--1.86 days after the burst (Sako \& Harrison 2002), can be used to 
derive the broadband optical-to-X-ray slope
$\beta_{ox} = -1.05$. Remarkably, this is similar to the X-ray spectral
index itself, $\beta_{x} \approx  -1.1$ $\pm$ 0.1.
However, a smooth extrapolation through the $BVRI$
photometric points yields $\beta_{o}$ $\approx -1.29$ and an even steeper
slope, $\beta_{o} \approx -1.66$, using the full
range of the LRIS spectral continuum. Although there is no
significant excess absorption in the X-ray afterglow spectrum (Sako
\& Harrison 2002), this type of discrepancy is common
in afterglow spectra and is normally
understood as requiring additional dereddening of the optical spectrum 
to account for local extinction in the host galaxy 
(\eg~Mirabal et al. 2002b). Alternatively the broadband spectrum 
can be described as having an X-ray excess due to inverse-Compton 
scattering (Sari \& Esin 2001).

The temporal decay described thus far is consistent with the predicted
adiabatic evolution of
a jet-like afterglow (Rhoads 1999). A gradual steepening of the optical
decay is expected when the jet angle begins to spread into a larger 
angle. Under the assumption
that the GRB is collimated initially, we estimate a half-opening angle 
of the jet~~$\theta_0 \approx 11^{\circ}\!n^{1/8}$ 
(Sari, Piran, \& Halpern 1999) for an
isotropic energy E$_{iso}\approx 5.6 \times 10^{52}$ ergs 
(Malesani et al. 2002). For frequencies 
$\nu$ $<$ $\nu_{c}$, where $\nu_{c}$ is the ``cooling frequency'' at which
the electron energy loss time scale is equal to the age of the shock,
the assumption of a synchrotron model in an uniform-density 
medium predicts $\alpha$ = (3/2)$\beta$ = $-3(p-1)/4$. Here $p$ is the index
of the power-law electron energy distribution.
For $\alpha_{o} = -0.72$, this implies 
$\beta_{o} = -0.48$ and $p$ = 1.96, which 
is consistent with the
optical data only 
if extinction at the host galaxy is significant (Holland et al. 2003). 

On the other hand, a model in which the afterglow expands into a 
pre-existing wind medium of density~$n \propto r^{-2}$ can reproduce
the slow decay at early times followed by
steepening caused by the synchrotron minimum characteristic 
frequency $\nu_{m}$ passing through the optical band 
(Li \& Chevalier 2003). The decay can be described by
$\alpha$ = $-(3p - 2)/4$ = $(3\beta + 1)/2$ for
$\nu$ $<$ $\nu_{c}$ (Chevalier \& Li 2000). 
A fit in the wind scenario 
yields $\alpha = -0.72$, with a steeper index
$\beta = -0.81$ and $p$ = 1.63. Although an electron 
index $p <$ 2 seems rather hard for a power-law
electron energy distribution, this type of electron distribution has been
encountered in other GRB afterglows 
(\eg~Panaitescu \& Kumar 2002). It is important
to note that a wind-like behavior seems to be supported by the
radio and X-ray observations assuming 
$\alpha = -1.0$ and $p$ = 2.1 (Li \& Chevalier 2003).  
It is difficult to determine a definite value for $\alpha$ because
of the ubiquitous fluctuations in the early optical light curve. 
The fact that
the broadband wind-interaction model provides a reasonable fit to the
early temporal decay $\alpha$, as well as to the spectral index 
$\beta$ without substantial
reddening, makes this model attractive
for a circumstellar medium with stellar-like density $n \propto$
$r^{-2}$. 
 
\subsection{Absorption System Identifications and Line Variability}

We used the full-range optical continuum of the GRB 021004 afterglow 
to derive  a function 
of the form 
$F_{\nu}$ $\propto$ $\nu^{\beta}$ with $\beta = -1.66$ $\pm$ 0.26, in
agreement
with the value reported by Matheson et al. (2003). As pointed out by these
authors, a shallower power-law index results from 
fitting only the red end of the spectrum. 
Three absorption systems are spectroscopically identified along the 
blue continuum  
at $z_{1}=1.380$, $z_{2}=1.602$, and 
$z_{3}=2.328$ that have been independently confirmed 
(\eg~Chornock \& Filippenko 2002; 
Salamanca et al. 2002; Matheson et al. 2003). 
In addition the spectrum reveals 
three distinct blueshifted absorbers at  
$z_{3A}$=2.323, $z_{3B}$=2.317,  and  $z_{3C}$=2.293 
within 3,155 km~s$^{-1}$ of the Lyman-$\alpha$ emission-line
redshift of the $z_{3}=2.328$ system (Chornock \& Filippenko 2002; 
Salamanca et al. 2002; Savaglio et al. 2002).

Figures 6 and  7~~show the normalized LRIS spectrum 
including emission and absorption-line systems, as well as
identified blueshifted absorbers. Table 2 lists the
line identifications including
vacuum wavelengths, 
observed wavelengths, redshift,
oscillator strengths $f_{ij}$, 
equivalent widths ($W_{\lambda}$) in the rest frame, and 
error estimates on the equivalent widths.
In order to compute the errors on the equivalent width for each line 
we used the IRAF {\it splot} task, which allows error estimates based on a 
Poisson model for the noise. For blended lines, IRAF 
{\it splot} fits and deblends each line separately using 
predetermined line profiles. Error estimates for blended lines 
are computed directly in {\it splot} by running a number of Monte Carlo 
simulations based on preset instrumental parameters.

There has been a recent suggestion of additional 
Lyman-$\alpha$ blueshifted absorbers located
at 27,000 and 31,000 km~s$^{-1}$ from the host galaxy (Wang et al. 2003).
Lines consistent with the reported positions are present in the LRIS 
spectrum; however, we believe that the proposed identifications are not
straightforward. Apart from being structured at 
the LRIS resolution, 
the lines lack matching C IV or Si IV blueshifted absorbers at the 
proposed velocities. 
An alternative identification is also plausible if the lines arise 
from Mg II doublets in systems 
located at redshifts $z$ $\approx$ 0.293 and 0.313,
respectively. However, the line ratios are inconsistent with this 
interpretation unless the lines are strongly saturated. Given the uncertainty
surrounding the nature of these lines, for the remainder of this work we will
characterize them as unidentified and will refrain from including them
in the analysis. We suspect that high-resolution spectroscopy of the optical
afterglow of GRB 021004 obtained by other groups (\eg~Salamanca et al. 2002) 
might provide more clues about these lines.

The prominence of the Lyman-$\alpha$ line emission and the presence 
of Al II (1670.79 \AA) in 
absorption at the same redshift as the Lyman-$\alpha$
emission, $z_{3}$ = 2.328,  confirms the highest 
system as the host galaxy of 
GRB 021004. The host galaxy is an active starburst galaxy with 
SFR $\approx$15 \Msol~yr$^{-1}$ (Djorgovski
et al. 2002). The detection of a 
lone low-ionization absorption line (Al II) at $z_{3}$ = 2.328 
seems plausible because of its large 
oscillator strength, $f_{Al~II}$ = 1.83. All other absorption lines 
(\eg~Lyman series, C IV, and Si IV) have velocity components blueshifted with 
respect to $z_{3}$. These components
are crucial to the analysis since intrinsic blueshifted absorbers located  
physically near the burst should be sensitive
to time-dependent photoionization due to the decaying GRB photoionizing
flux (Mirabal et al. 2002b; Perna \& Lazzati 2002). Although many of the 
absorption lines are not fully resolved, the C IV and Si IV doublet ratios 
suggest that the $z_{3C}$ absorber is not strongly saturated. 
Other absorbers are blended, but do not show flat profiles reaching zero 
intensity which are a distinct indication of strong saturation. 

Direct comparison of the equivalent-width measurements 
presented in this work with  
published results (M$\o$ller et al. 2002; Matheson et al. 2003)
show no definite  evidence for time-dependent
absorption-line variability on timescales of hours 
to days after the 
burst. In addition, there are no strong observable signatures of immediate 
production of vibrationally excited H$_{2}$
levels in the region 912 \AA\ $\leq$ $\lambda_{\rm rest}$ 
$\leq$  1650 \AA~(Draine 2000), and reradiated fluorescent 
emission in a similar range. The recent report of spectropolarimetric
variations seen across some Lyman-$\alpha$ absorption features, 
and the increasing polarization near the blue continuum of
the GRB 021004 afterglow (Wang et al. 2003), 
are reminiscent of the effects reported 
in broad absorption-line QSOs (Goodrich \& Miller 1995). If real, 
the spectropolarimetric results
would favor the proximity of the absorbers to the burst. This possibility
may be reinforced by the suggestion of  
a ``line-locking'' effect (\eg~Scargle 1973) in
the C~IV doublet (Savaglio et al. 2002). 

\subsection{Abundance Analysis}

In order to derive the abundances 
of the identified absorbers, we estimated the 
column density $N_{j}$ for each identified line $j$ following the 
linear part of the curve of growth (Spitzer 1978) written in the form
\begin{equation}
N_{j}({\rm cm}^{-2})= 1.13 \times 10^{17}\frac{W_{\lambda}({\rm m\AA})}{f_{ij}\lambda^{2}({\rm \AA})}.
\end{equation}
\noindent
The resulting column densities derived for single absorption lines 
are listed in Table 3. A visual inspection of the lines 
does not reveal strongly saturated 
profiles; however, most lines are not fully resolved. 
Comparison with Table 3 shows that the hydrogen column
densities inferred from Lyman-$\alpha$ are less than those inferred
from Lyman-$\beta$. This might be the result of line blending or simply 
implies that Lyman-$\alpha$ is somewhat saturated.

Resulting total ionic concentrations are given in Table 4.  
In order
to determine the total abundances of each element, we assumed the observed 
ionic concentrations and upper limits for various states of ionization in
the spectral range. 
Therefore, the abundances obtained are an
underestimate of true abundances since ionic abundances of other species
are required. However, we justify this simplified scheme by pointing out
the approximate coincidence in ionization potential of various species
(Si, C, N) and the detection of the dominant ions for each element. 
Particularly interesting are the measurements of C and 
Si since they exhibit enhanced abundances 
compared to solar abundances (Anders \& Grevesse 1989). This is
discussed further in \S 3.4.  The largest uncertainty is that of 
oxygen due to the large ionization potential of its high-ionization states.
Since no O VI was detected, it
is impossible to predict what ionization states of oxygen should have
been present along this line of sight (Spitzer 1996; Savage et al. 2000). 

In general, ionization effects depend on the 
conditions of the environment. For this reason, in most elements, 
ionization processes are complex and layered around
the GRB host galaxy. Accordingly, the dominant sources
of error in the total abundances are the uncertainty in the temperature
of the medium and the errors in the measured equivalent 
widths (Savage \& Sembach 1996).
We note that the observed blueshifted absorbers range
in ionization from Lyman-$\alpha$ $\lambda$1215.67 \AA~to 
C IV $\lambda\lambda$1548.20 \AA, 1550.77 \AA. 
The presence of Lyman-$\alpha$ in absorption 
indicates a low-ionization gas component that cannot survive 
in the highly ionized C IV/Si IV region unless hydrogen is shielded
from external photoionization or is dense enough to recombine. 
One plausible scenario is that 
we are probing a shielded low-ionization region that  
has been enriched by physically adjacent C~IV and Si~IV.

\subsection{Kinematics and Abundances of the Blueshifted Absorbers}

The next step in our analysis is to explore a connection between 
the chemical abundances
and the physical mechanism responsible for accelerating the 
blueshifted absorbers. Starting with the hypothesis that the 
absorbers are intrinsic to the host galaxy, we 
recall that multiple blueshifted absorbers at similar velocities
have been detected in massive stellar winds  
(Abbott \& Conti 1987), as well as in QSO absorption-line 
systems (Anderson et al. 1987). The former are understood to be
driven by the pressure of the stellar radiation (Castor, Abbott, \& 
Klein 1975), while the latter are thought to arise either in 
chance intervening neighboring systems or as part of QSO gas outflows 
(Aldcroft, Bechtold, \& Foltz 1997). 
One important distinction in this instance 
is the absence of any obvious spectroscopic evidence 
for an active QSO associated with the host galaxy of GRB 021004. 
We shall consider the likelihood 
of chance intervening systems in \S 6.

Having argued against a QSO-related origin, we focus on the 
possibility of a massive stellar wind around the GRB progenitor. 
This scenario is highly relevant in connection to  
massive-star progenitors in GRB models (Woosley 1993).
Current stellar models predict that a massive star 
loses most of its original hydrogen envelope via stellar winds 
exposing elements like carbon, nitrogen, and oxygen (Abbott \& Conti 1987). 
This stage marks the 
beginning of the Wolf-Rayet phase. According to the observed 
chemical composition, Wolf-Rayet stars are
classified into subtypes WC, WN, and 
WO (Crowther, De Marco, \& Barlow 1998). 
For instance, in a few  WN stars, hydrogen appears to be present along with
helium and nitrogen lines, while the   
majority of WC and WO Wolf-Rayet stars display hydrogen-free spectra. 
The notable absence of helium, nitrogen and oxygen in the spectrum
of GRB 021004 seemingly rules out a straightforward connection 
with WN and WO subtype stellar winds. 
A bigger burden for a smooth stellar wind scenario results from the  
uncomfortable task of placing sufficient low-ionization species in the 
same region as highly ionized species like C IV and S IV 
once the photoionization front
from the GRB has made its way through the wind. This is because 
most species in a stellar wind, following 
a $n \propto r^{-2}$ profile, 
are completely photoionized within a few parsecs of the
GRB almost instantly (see also Lazzati et al. 2002).

Based on the previous reasoning, it appears unlikely that 
the observed absorbers  
are produced directly within a smooth stellar wind. However, 
we have yet to consider the interaction of 
a stellar wind with its 
neighboring ISM and material shed during previous stellar phases. 
A massive stellar wind 
carries not only mass but kinetic
energy that produces shocks in the wind-ISM interaction (Castor, McCray, \& 
Weaver 1975; Ramirez-Ruiz et al. 2001). The interaction leads naturally to 
the formation of 
overdense shells or shell nebulae along the
wind profile, as seen in numerous examples (\eg~Moore, Hester, 
\& Scowen 2000). These observations suggest that shell nebulae are common
around Wolf-Rayet stars. Indeed, narrow-band surveys indicate that 
shell nebulae are present around 35$\%$
of all Wolf-Rayet stars (\eg~Marston 1997). 
A study of the optical morphologies of shell nebulae shows distinctions
between different stages of formation and physical conditions
of their interior (Chu 1991).

Apart from providing a complex circumstellar 
environment, a shell nebula configuration
enables natural mixing of low-ionization hydrogen species 
from the ISM and prior main sequence/supergiant phases, 
with high-ionization C~IV and Si~IV from an adjacent 
Wolf-Rayet wind. For instance, 
nebular structures observed around the explosion site of SN 1987A 
(Panagia et al. 1996 and references therein) are believed 
to have been enriched by the progenitor material prior to the 
explosion (Wang 1991).
A number of spectroscopic observations confirm
that shell nebulae around Wolf-Rayet stars 
are mainly material from the
massive star rather than the ISM (Parker 1978; Kwitter 1984).
The absence of strong nitrogen and oxygen lines, and the
presence of C IV and Si IV in the GRB 021004 afterglow spectrum are consistent
with a WCL Wolf-Rayet 
star (Mirabal et al. 2002a), in which the bulk of the wind 
has a composition
characteristic of He-burning and $\alpha$-capture
products (Crowther, De Marco, \& Barlow 1998).
This line of 
argument is thus far consistent with a shell nebula observed
as chemical enrichment in the blueshifted absorbers, but let us 
explore its kinematic evolution.

\section{The Expansion of a Shell Nebula}

The free expansion of a massive stellar 
wind is thought to end when 
the mass of the swept-up  shell nebula 
is comparable to the mass driven by the wind (Castor,
McCray, \& Weaver 1975). Figure 8 shows the 
theoretical structure of a stellar wind bubble 
and shell nebula 
formed at the termination of a free-expanding wind. 
The swept-up shell nebula mass becomes equal to mass driven by 
the wind at a time $\tau$ set by  
\begin{equation}
\tau=\sqrt{3\dot{M} \over 4 \pi v_{t}^3 n m_{\rm
p} \mu } \approx 300~{\rm yr},
\end{equation}
\noindent
for a typical mass-loss rate~~$\dot{M}=10^{-5}~{\rm M}_{\odot}$~yr$^{-1}$, 
density of the surrounding medium $n$ = 1 cm$^{-3}$, and terminal velocity 
$v_{t}$ = 1000 km~s$^{-1}$. The mass conservation relation
implies that during this time a stellar wind 
moving at $v_{t}$ = 1000 km~s$^{-1}$ has reached a radius $R_{\rm s}$ 
given by
\begin{equation}
R_{\rm s} = v_{t} \tau \approx 0.3~{\rm pc}.
\end{equation}
\noindent
This radius $R_{\rm s}$ is 
in agreement with the modeling of Wolf-Rayet stars using
detailed stellar tracks (Ramirez-Ruiz et al. 2001).  
After the swept-up shell nebula is 
formed, it proceeds to expand adiabatically  
because the pressure of the hot gas inside the wind bubble 
is higher than the circumwind environment (Castor,
McCray, \& Weaver 1975). 
As it expands, a low-ionization swept-up shell nebula formed 
around a massive-star bubble   
will be gradually enriched and 
fragmented as it is subject to Rayleigh-Taylor 
and  Vishniac
instabilities (Ryu \& Vishniac 1988; Garc{\ia}a-Segura 
\& Mac Low 1995a,b). The onset of instabilities 
would explain naturally the presence of multiple 
dense-shell fragments along this
line sight that could give rise to the
individual blueshifted absorbers observed in the spectrum of the GRB 021004
afterglow. 

The expansion of the shell nebula in the adiabatic phase 
can be described by the momentum equation, or
\begin{equation}
{d \over dt}[{M_{\rm s}(t)v(t)}]=4\pi R_{\rm s}^{2} P_{w},
\end{equation}
where $M_{\rm s}(t)$ is the mass of the swept-up shell nebula, $v(t)$ is the
rate of expansion of the bubble, and
$P_{w}$ is the internal pressure caused by the wind.
In the adiabatic regime, the internal pressure due to
the wind can be written as $P_{w}$ = $L_{w}t/(2\pi R_{\rm s}^{3})$,
where $L_{w}$ is the wind luminosity. 

Using
this expression for the internal pressure gives
\begin{equation}
{R \over t} {d \over dt} \left(R^{3} {d \over dt} R\right) = {3 L_{w} \over
2 \pi n m_{p}},
\end{equation} 
where we have 
used $v(t)=dR/dt$ and $M_{\rm s}(t) = (4\pi/3) R_{\rm s}^{3} n m_{p}$. The 
expression has a solution of the form

\begin{equation}
R_{\rm s}(t) = \left({25 L_{w} \over 14 \pi n m_{p}}\right)^{1/5} t^{3/5},
\end{equation}
\noindent
which can be rewritten as 
\begin{equation}
R_{\rm s}(t) = 33 \left({L_{36} \over n_{0}}\right)^{1/5} t_{6}^{3/5} {\rm pc},
\end{equation}
with $L_{36}$ in units of $10^{36}$ erg~s$^{-1}$, 
$n_{0}$ in units of cm$^{-3}$, and $t_{6}$ in units of $10^{6}$~~yrs. 
The
velocity of expansion of the shell nebula is given in the same terms by
\begin{equation}
v(t) = 19.8 \left({L_{36} \over n_{0}}\right)^{1/5} 
t_{6}^{-2/5}~{\rm km~s^{-1}}.
\end{equation}

A key result here is that over the duration of the Wolf-Rayet phase,   
shell nebulae can reach radii of order $R_{\rm s} \approx$ 10 pc and  
expansion velocities $v \approx$ 
40 km~s$^{-1}$. Evidently the derived expansion velocity of
a swept-up shell nebula is nowhere near the observed $\sim$ 450, 
$\sim$ 990, 
and  $\sim$ 3,155 km~s$^{-1}$ blueshifted absorbers. 
If instead of an energy-conserving expansion, we invoke large 
radiation losses and assume that the wind bubble is undergoing 
momentum conservation and hence expanding 
as $R_{\rm s}(t) \propto t^{1/2}$ (Steigman, Strittmatter, \& Williams 1975),
the approximation yields a radius and expansion velocity 
similar to the energy-conserving 
solution and is still inconsistent with the observed velocities. 
We call this inconsistency with the  blueshifted absorbers
the {\it kinematic problem}.

\section{Radiative Acceleration of a Shell Nebula}

Faced with a theoretical expansion velocity 
much too slow to explain the blueshifted absorbers, 
we reexamined the
velocity profiles that we obtained for Lyman-$\alpha$, Lyman-$\beta$, 
C~IV, and Si~IV. If the blueshifted components are associated with
the host galaxy of GRB 021004, these must 
originate in an expanding outflow or alternatively might have been 
accelerated radiatively by the GRB. The absence of noticeable absorption-line 
variability and 
deceleration in the absorber velocities could be an argument  
against an expanding outflow near the
GRB afterglow. An outflow leading the GRB afterglow 
would most likely be subject to rapid photoionization and
even disappear as the shock overruns it. 
An alternative model assumes that radiative acceleration 
by the GRB afterglow plays a crucial role in the kinematics
of the wind bubble and shell nebula surrounding a 
Wolf-Rayet progenitor. The advantage here is that 
radiative acceleration provides more flexibility in 
the discreteness and velocity structure of
the blueshifted absorbers. Radiative acceleration effects in absorption
could also lead to ``line-locking'' as suggested by high-resolution
spectroscopy (Savaglio et al. 2002). 

We can directly model the radiative history of a wind-bubble/shell-nebula 
system by using photoionization models
with a fixed prescription for the density profile. For these particular 
simulations, we have 
used the photoionization code IONIZEIT (Mirabal et al. 2002b),
which includes time-dependent 
photoionization processes taking place under 
a predetermined GRB afterglow ionizing flux. 
Recombination processes are not included since the densities to be 
considered are not sufficiently high to produce a recombination timescale
comparable to the duration of the GRB afterglow. 
This is a major assumption since
the densities at which the recombination timescales become comparable 
to the duration of the bright phase of the 
afterglow, $10^{10}-10^{12}$ cm$^{-3}$ (Perna \&
Loeb 1998), are still allowed on the basis of  
high-resolution X-ray spectra of GRB afterglows 
(Mirabal, Paerels, \& Halpern 2003). Moreover, observations of water masers in
circumstellar envelopes suggest 
densities of $\sim 5 \times 10^{9}$ cm$^{-3}$ within discrete clumps
(Richards, Yates, \& Cohen 1998),  
which would significantly reduce the recombination timescale within 
overdensities. 

In each case
the densities and
physical regions are chosen to match the observed column densities (Mirabal
et al. 2002b). The models used here 
include elemental abundances of H, He, C, and Si.  
The input flux $F_{\nu'}(r,t')$ was approximated from the broadband
observations of GRB 021004. The functional form for the flux 
$F_{\nu'}(r,t')$ has two components to accommodate the observed
``rise'' in the optical light-curve at~$t_{\rm rise} \approx$0.08 days (Fox
et al. 2003). So, for $t \leq$ 0.08 day, 

\begin{equation} 
F_{\nu'}(r_0,t')= 2.21 \times 10^{-26}   
\left({\nu' \over 4.55 \times 10^{14}(1+z) {\rm Hz}}\right)^{-1.05}
\left({d^{2}\over (1 + z)r_{0}^{2}}\right) 
\left({t'(1+z)\over 0.0066\, {\rm day}}\right)^{-0.8}
{\rm {ergs~cm^{-2}~s^{-1}~Hz^{-1}}};
\end{equation}

\noindent
otherwise 

\begin{equation} 
F_{\nu'}(r_0,t')= 4.66 \times 10^{-28} 
\left({\nu' \over 4.55 \times 10^{14}(1+z) {\rm Hz}}\right)^{-1.05}
\left({d^{2}\over (1 + z)r_{0}^{2}}\right) 
\left({t'(1+z)\over 1.37\, {\rm day}}\right)^{-0.72}
{\rm {ergs~cm^{-2}~s^{-1}~Hz^{-1}}},
\end{equation}

\noindent
where $d$ is the luminosity distance to the burst at 
$z=2.328$ (assuming $H_0 \simeq 65$ km~s$^{-1}$~Mpc$^{-1}$, $\Omega_m \simeq 
0.3$, $\Omega_{\Lambda} \simeq 0.7$), and $r_{0}$ is the inner radius of
the photoionized region set by the shock evolution 
$r_{0}=2.85 \times 10^{16}t_{days}^{1/2}$ cm (Chevalier \& Li 2000). 
The simulations also take into account 
the effect of synchrotron self-absorption during the initial seconds 
(Piran 1999). 

Throughout we adopted
a standard $n \propto r^{-2}$ scaling and shell-nebula fragments
with a density 
$n_{\rm s}\approx 10^{3}-10^{6}$ cm$^{-3}$, motivated by observations 
(Moore, Hester, \& Scowen 2000). 
Initially, we  
considered the simplest smooth wind model for the density profile with
overdense shell-nebula fragments superposed. 
The parameters of the IONIZEIT models were then varied to maximize 
the agreement with the observed blueshifted absorbers.
In order to avoid overionization, the absorbers must be dense with the
appropriate filling factor 
or alternatively the shell-nebulae fragments must be shielded from
the GRB emission by attenuating 
optically thick material at the base of the wind bubble. 
Satisfactory photoionization models require the shell-nebula fragments 
to be placed at a distance of at least 
$R_{\rm s} \gax$ 0.3 pc to reproduce the non-detection of
absorption-line variability in GRB 021004. 
Using the derived column densities 
and assuming that we are looking at a typical line of sight,
we can estimate the physical mass of each fragment $\Delta M$, where 
\begin{equation}
\Delta M= 4\pi R_{\rm s}^{2} \Delta R n_{\rm s} m_{p} \mu \gax 10^{-4}~{\rm M_{\odot}}.
\end{equation}   
\noindent
In the case 
$R_{\rm s} \gax$ 90 pc, this implies 
$\Delta M \gax$ 10 M$_{\odot}$ which sets a
tentative upper limit on the shell-nebula radius simply 
based on the mass-loss rate. 

With these initial constraints, we proceeded to use 
the IONIZEIT code to calculate the radiative momentum 
acquired within individual shell-nebula fragments.
The fine-tuning for any configuration derives from the balance required 
to prevent extreme overionization of the blueshifted absorbers 
and still be efficient for acceleration mechanisms. 
In particular, 
the radiative acceleration $g(r,t)$ as a function of time can be expressed as
\begin{equation}
g(r,t) = {\kappa(r,t) L(t) \over 4 \pi r^{2} c},
\end{equation}
where $L(t)$ corresponds to the total luminosity and
$\kappa(r,t)$ represents the opacity at a distance
$r$. The radiative flux as a function of time can be 
estimated directly within each shell-nebula fragment by following the 
prescription in Mirabal et al. (2002b):
\begin{equation}
F_{\nu}(r_{i+1},t) = F_{\nu}(r_{i},t) e^{-\tau_{\nu,i}}
\left({r_{i} \over r_{i+1}}\right)^{2},
\end{equation}
where $\tau_{\nu,i}$ stands for the photoionization optical depth 
which is estimated within each shell-nebula fragment $i$. 
The product of the radiative acceleration and the interval 
between time steps $\Delta t$ yields the total velocity acquired 
by a shell-nebula fragment as a function of time, 
\begin{equation}
v(t) = v_{o} + \sum_{r,t} g(r,t) \Delta t,
\end{equation}
where $v_{o}$ is the initial velocity in the shell nebula. This calculation
assumes that the blueshifted absorbers are driven mainly
by bound-free 
absorption transferred to the shell nebula fragments. 
Additional mechanisms that can contribute to the radiation 
acceleration term are bound-bound processes, 
free electron scattering, and line driving.   
Generally, spectral lines can play an important factor in  
enhancing the electron scattering coefficient (Castor, Abbott, \& 
Klein 1975; Gayley 1995). However, the available 
time for scattering after the GRB 
is much shorter than in long-lived stellar winds
or active galactic nuclei where line driving might be most 
efficient (\eg~Proga, Stone, \& Kallman 2000). A full two-dimensional, 
time-dependent
simulation of a radiation-driven wind around a GRB is
imperative to determine the contribution from different 
mechanisms.

Figure 9 illustrates the total velocity acquired by a fragmented shell nebula 
as a function of time. The model assumes that the shell nebula
is distributed over a thick annulus located $\gax$ 0.3 pc from the GRB and 
that the fragments are overdense at 0.3 pc, 0.54 pc, and 0.8 pc.  
Clearly, the radiative acceleration model shown in Figure 9 
reproduces the total velocity required to accelerate
individual blueshifted absorbers to the observed velocities.
These results are in agreement with the discussion by Schaefer et al. (2003).
In order to reach the observed velocities and avoid major absorption-line 
variability, the bulk of the radiative acceleration needs to take place
during the early stages of the afterglow, which is consistent with the model.
The faster-moving fragments will get 
impacted by a larger flux and acquire more radiative acceleration.
The slower fragments can be explained
reasonably if they are more distant or less opaque 
than the fragment closest to
the GRB. In general, shell nebulae can present low 
opacities to radiative flux. This seems to
be confirmed by observations of the NGC 6888 nebula 
where only 2$\%$ of the ionizing photons are thought to be 
processed within the shell nebula (Moore, Hester, \& Scowen 2000). 
Alternatively, the slower fragments might have been subject to 
deceleration as these encountered the surrounding medium. 
Although our simulations can reproduce the velocities 
of the absorbers, we cannot rule out
that the absorbers are very distant and 
completely unrelated to the GRB event. However,
the spectropolarimetric results (Wang et al. 2003) hint at an intrinsic
origin for the absorbers.  

For simplicity, 
processes such as multiple photon scatterings, 
density gradients within each fragment, and 
dust destruction/acceleration have been ignored but warrant
consideration in more detailed modeling of radiative acceleration processes
around GRBs.  
Because we were denied access to the true broadband GRB photoionizing flux
at early times, the models described thus far should be considered
tentative. While it can be argued that 
the actual GRB photoionizing flux, density structure, and 
opacity within the shell-nebula fragments could be quite different, 
we believe that variations about the initial estimates can be accommodated by 
modifying the placement and density structure within each shell-nebula 
fragment without
altering our main conceptualization. It is important to note that 
observed shell nebulae span diameters ranging from 0.3 pc to 180 pc 
(Marston 1997; Chu, Weis, \& Garnett 1999) and that  
only about 35$\%$ of all Wolf-Rayet stars seem to be surrounded by 
overdense 
shell nebulae (Marston 1997). Furthermore, shell nebulae 
typically display intrinsic
expansion velocities $v \approx 40$ km~s$^{-1}$ that can only be resolved
with high-resolution spectroscopy.  
Taken together, these facts imply that 
shell nebulae around GRBs might have been missed in the past 
either because they were absent, too slow, or completely 
photoionized by the GRB emission. Another important factor is the 
morphology of shell nebulae that might have a decisive effect 
in the angular geometry of the absorbing material (Chu et al. 1991). 
GRB 021004 could be a
fortunate instance where the shell nebula around a GRB progenitor 
was located at an ideal distance from the GRB to
avoid complete photoionization and simply acquire sufficient  
radiative acceleration to produce resolved individual blueshifted 
absorbers. 

\section{Alternative Explanations for the Blueshifted Absorbers}

\subsection{Supernova Remnant}

Having made an argument for 
accelerated shell-nebula fragments to explain the abundances and kinematics of
the blueshifted absorbers, we now evaluate whether the observations can
still be compatible with a different origin.  
Of the numerous models 
for GRB progenitors, the supranova model (Vietri \& Stella 1998)
and the magnetar models (Wheeler, Meier, \& Wilson 2002) predict a 
possible association with a supernova remnant (SNR) that would 
already be in place prior to the GRB onset.
This possibility has been raised to explain the blueshifted
absorbers in the GRB 021004 afterglow spectrum (Wang et al. 2003) and its 
deviations about the light curve (Lazzati et al. 2002). 

Assuming that the observed velocities
reflect the mechanical momentum acquired during the 
free expansion of the SNR together with 
the distance constraint obtained 
from the photoionization simulations ($R_{\rm s} \gax$ 0.3 pc) 
yields a minimum 
age for the remnant $t_{SNR}$ $\gax$ 100 yrs.
 The estimated age, $t_{SNR}$, appears high relative 
to simulations of neutron stars which show major 
difficulties  maintaining differential rotation in neutron stars 
for more than a few minutes (Shapiro 2000). However, $t_{SNR}$ is
 still barely consistent with the analytical  
supranova model which assumes magnetic fields of $\approx 10^{12}-10^{13}$ G,
and a SNR age of a few weeks to several years ($\sim$ 100 yrs) (Vietri \&
Stella 1999). Possibly a bigger difficulty
facing the SNR scenario is the absence
of strong blueshifted Al, Fe, and O absorbers 
that should be evident in the
remnant of a core-collapse SN (Hughes et al. 2000; Patat et al. 2001). 

Considering that the observed abundances 
are those around a GRB progenitor, then a massive star that is
part of a binary system embedded 
within the old SNR of its companion is also a possibility (Fryer et al. 
2002). In that scenario, the hydrogen envelope of the actual GRB 
progenitor might 
have been lost via mass transfer to a companion that exploded as a 
SN following mass 
transfer. Only after removal via mass transfer 
of the shear created by a hydrogen envelope, 
the actual GRB progenitor might have retained sufficient angular momentum 
($j \gax$ $10^{16}$ cm$^{2}$~s$^{-1}$) to produce a 
collapsar (MacFadyen \& Woosley 1999). 
Apart from envelope stripping, an additional advantage of a binary
system is the collision of stellar winds that can produce turbulence 
(Kallrath 1991; Stevens, Blondin, \& Pollock 1992) and
could account for the clumpy structure observed in the optical 
decay. This latter 
scenario is still consistent with a Wolf-Rayet star GRB progenitor. 

\subsection{QSO Absorption-Line Systems}

QSO absorption-line systems provide a more obvious connection
to blueshifted absorbers. There are numerous QSO observations  
displaying prominent high-velocity 
blueshifted absorbers (\eg~Weymann et al. 1979; Anderson et al. 1987). 
These narrow lines are thought to form either in ejecta or infall near the QSO
or in intervening
systems that coincidentally fall along the line of sight to the QSO. 
An examination of the GRB 021004 afterglow spectrum reveals no definite 
evidence that the host galaxy is an active 
QSO, hence a connection with intrinsic QSO  
gas outflows is not implied. Nevertheless, we cannot rule out
the possibility that a QSO accelerated the absorbers and  
became dormant after a duty cycle of  
$\sim 10^{7}$ yrs (Wyithe \& Loeb
2002). The scenario
does require that the QSO outflow took place 
nearly aligned with the line of sight to the GRB, which 
seems highly improbable.

\subsection{Supershells and Superwinds}

The inferred SFR $\approx$ 15 \Msol~yr$^{-1}$ 
for the host galaxy of GRB 021004 (Djorgovski et al. 2002) 
is well above the average rate at that redshift.
Interestingly, a number of powerful extragalactic starbursts 
show emission-line outflows 
at velocities around $10^{2}-10^{3}$ km~s$^{-1}$ (Heckman, Armus, \& 
Miley 1990). 
The majority of these ``superwind'' measurements are
made from emission-line widths. In the case of GRB 021004, the blueshifted 
absorbers are resolved and span a larger velocity range than the wind 
velocity inferred from the Lyman-$\alpha$ emission-line 
profile. If a large-scale superwind venting into the halo of the host galaxy 
is responsible for the blueshifted absorbers, one might expect  
Al II from interstellar gas to be blueshifted with respect to the 
Lyman-$\alpha$ emission as part of the expanding outflow 
(Heckman et al. 2000).  This is not the case in the 
GRB 021004 afterglow spectrum 
(\S 2). A different possibility is a chance interception
of three local supershells associated with star-forming regions within
the host galaxy driven 
by SNe and stellar winds in starburst bubbles (Heiles 1979). 
In theory, the large SFR could lead naturally to multiple 
energetic OB associations ($\gax$ 1000 stars); however,
velocities $\geq$ 500 km~s$^{-1}$ are rarely observed in individual 
shells around our Galaxy (Heiles 1979).

\subsection{Outflowing Systems}

In addition to the well-established intrinsic absorbers, there is a
possible association with intervening gas extended 
over 3,155 km s$^{-1}$ and observed in projection along this line of sight. 
The system could be a very high-velocity analog of   
local outflowing systems (Savage et al. 2003). However, an extension 
of structure
over 3,155 km s$^{-1}$ in velocity space appears highly unlikely 
based on the observed velocity distribution through the Milky Way. 
Moreover, the host galaxy would
have to spill metals within the Lyman-$\alpha$ clouds to create 
the observed metal enrichment. Finally, a distant origin 
would be ruled out if the reported polarization changes across 
the Lyman-$\alpha$ absorption and continuum are intrinsic to the host 
galaxy (Wang et al. 2003).  

\section{Implications for the GRB Progenitors}

Even though we cannot yet rule out definitely some of the alternative 
explanations, it is apparent from the analysis that a shell nebula 
around a massive-star progenitor is likely to give rise to the blueshifted 
absorbers in the spectrum of the GRB 021004 afterglow.  
The large deviations in the 
optical decay of the GRB 021004 afterglow (see \S 3.1) are unusual and 
suggest that additional effects such as  
small-scale inhomogeneities in the circumburst medium (Wang \& Loeb 2000;
Mirabal et al. 2002a),
structure within a jet (Kumar \& Piran 2000), 
and/or ``refreshed'' collisions among separate shells of ejecta 
are taking place (Rees \& M\'esz\'aros 1998). 
Different groups have fitted the 
$R$-band data (Lazzati et al. 2002; Nakar et al. 2003), as well
as the broadband data (Heyl \& Perna 2003), to explore each possibility. 
Although several
models provide reasonable fits to the $R$-band data, the broadband
modeling finds that a clumpy medium produced by density fluctuations
provides a more reasonable fit to the data (Heyl \& Perna 2003). 
The interpretation of density fluctuations in the GRB 021004 circumburst 
medium is entirely consistent with the predicted density bumps
that arise when stellar winds sweep up the ISM or the material shed by the
star in previous stages of evolution (Mirabal et al. 2002a;
Ramirez-Ruiz et al. 2001). 
It is also possible that a cocoon from a progenitor stellar envelope
can be displaced along the direction of the GRB relativistic jet 
(Ramirez-Ruiz, Celotti, \& Rees 2002).
A number of observations of Wolf-Rayet
stars confirm that stellar winds are indeed not homogeneous but rather
clumpy (Nugis, Crowther, \& Willis 1998; L\'epine et al. 1999). 

Upon examination of Figure 3, it is clear that   
the OT also 
exhibits a distinct color evolution over time (Bersier et al. 2003;
Heyl \& Perna 2003). On its way to the Wolf-Rayet phase, a main-sequence
star is thought to evolve into a supergiant phase (Abbott \& Conti 1987).
The mass loss in the supergiant phase leads to the formation of a dense
supergiant material shell. After entering its Wolf-Rayet phase, the
Wolf-Rayet wind slowly starts sweeping the supergiant material, eventually
overtaking the main-sequence material 
from the star. The streaming of winds, and wind collisions 
taking place throughout the mass-loss history
of the star, result in a complex morphology that might lead to distinct 
color changes and a spectrum redder than the typical 
synchrotron spectrum (Ramirez-Ruiz et al. 2001) as seen in Figure 3, 
especially if these are dusty winds accelerated by the 
stellar luminosity. We postulate that if the color changes are
external to the afterglow/jet evolution, the changes 
might be intrinsically related 
to the mass-loss history and dust patterns within a 
massive stellar wind 
(Garc{\ia}a-Segura \& Mac Low 1995a,b). Two-dimensional gasdynamical  
wind simulations including dust are necessary to explore this possibility.

The suggestion of a fragmented shell nebula around the GRB 021004 
progenitor accompanied by a clumpy wind medium 
meets partially the conditions required
by the collapsar model (Woosley 1993). It is associated with a massive
star and a star-forming region (MacFadyen \& Woosley 1999). 
The main theoretical difficulty with the collapsar model has been the 
requirement for retaining sufficient  
angular momentum (MacFadyen, Woosley, \& Heger 2001). Possible solutions 
include  metal-deficient stars and/or Wolf-Rayet stars that have lost 
most of their envelope through an efficient 
progenitor wind or to a binary companion
(MacFadyen \& Woosley 1999). These solutions remove the torques
induced by an outer envelope and conserve adequate rotation. The 
interpretation of an enriched shell-nebula around the GRB 021004
progenitor hints at the possibility that a massive-star GRB progenitor
might have lost most of its envelope prior to collapse. 
If this were the case, a stripped core 
would ease conservation of angular momentum 
requirements prior to iron-core collapse 
and support a connection with the collapsar GRB model.   
Unfortunately, due to our limited access to a single line of 
sight towards the GRB, there is little information about the 
three-dimensional geometry and evolution of the collapse. Therefore, 
it is crucial to complement time-variability studies 
with contemporary polarization measurements that might provide
information about the evolution of the jet (Sari 1999).

\section{Conclusions and Future Work}

The presence of blueshifted absorbers in the spectrum
of the GRB 021004 afterglow presents possible evidence
for a fragmented shell nebula located $\gax$ 0.3 pc from the GRB site 
that has been radiatively accelerated by the GRB afterglow 
emission. While at this stage 
we cannot rule out an origin related to a dormant
QSO, large-scale superwinds, 
or an old supernova remnant,  these alternative
explanations present some problems. 
The mass-loss process in certain 
massive stars might conserve sufficient angular momentum to
induce an efficient iron-core collapse or collapsar. If this interpretation
is correct, the observational data on GRB 021004 might be the first direct
evidence of a Wolf-Rayet star GRB progenitor.
Additional spectroscopy of high-ionization absorbers such as C IV, Si IV,
N V, and O VI along with associated low-ionization species 
will clarify this possibility, with the caveat that nearby 
shell nebulae might be rapidly photoionized by the GRB and 
that only 35$\%$ of all Wolf-Rayet stars show evidence of  
overdense shell nebulae. In this context, the advent of the {\it Swift}
 satellite (Gehrels 2000) should provide unique access  to
early multiwavelength observations of GRB afterglows that 
will be fundamental for determining the photoionization history and 
radiative acceleration  evolution of absorbers.

Interestingly, the inhomogeneities about the 
optical decay of the GRB 021004 afterglow imply  
that overdensities in a clumpy medium might be responsible for 
bumps in the OT decay. This finding motivates the need to model 
highly structured circumburst media beyond the simplest 
uniform and wind-like profiles.
It also calls for dedicated observatories and observers
to provide continuous coverage for a bigger sample of GRB afterglows.  
It is possible that overdensities   
might explain the presence of some late-time secondary peaks seen in other
GRBs (\eg~Bloom et al. 2002; Garnavich et al. 2003) if SN spectral 
signatures are
missing in the late-time spectrum. In fact, a 
consequence of the shell nebula model is that a rebrightening in the 
light curve should occur  
once the shock overruns the shell-nebula fragments 
(Ramirez-Ruiz et al. 2001). 
In addition, blueshifted absorbers from a shell nebula should disappear as
the shock reaches that point. 
Unfortunately, by the time this were to happen in the GRB 021004 afterglow 
decay ($\gax$ 1 yr after the burst), the light would be completely 
dominated by the 
host galaxy. Continued late-time photometry and spectroscopy 
is urged in order to search for this definite signature in other GRBs.
Finally, if some GRBs are produced by core-collapse in Wolf-Rayet stars, 
type Ib or Ic
supernovae might be a viable consequence after the violent event (Smartt 
et al. 2002). The recent discovery of SN 2003dh associated with GRB 030329
(Stanek et al. 2003; Chornock et al. 2003) could provide  further 
constraints on the nature of the
GRB progenitors and another link between Wolf-Rayet stars and GRBs.

\acknowledgments{ }
We would like to thank Sebastiano Novati and Vincenzo Cardone 
for obtaining observations at the MDM Observatory, and the Keck Observatory
staff for their assistance. We thank 
Jim Applegate, Orsola De Marco, and Mordecai-Mark Mac Low for useful 
conversations. We also acknowledge Eric Gotthelf for allowing us to use his 
Alpha computer. This material is based upon work supported by the 
National Science Foundation under Grants AST-0206051 to J. P. H. and 
AST-9987438 to A. V. F.

\clearpage

\begin{deluxetable}{lcrc}
\tabletypesize{\footnotesize}
\tablenum{1}
\tablewidth{0pt}
\tablecaption{Optical Photometry of GRB 021004}
\footnotesize
\tablehead
{Date (UT) & Filter   & 
Magnitude & Telescope} 
\startdata
2002 Oct 5.118 & $B$ & $19.95 \pm 0.10$ & MDM 1.3 m\\
2002 Oct 5.143 & $B$ & $19.90 \pm 0.10$ & MDM 1.3 m\\
2002 Oct 5.169 & $B$ & $20.09 \pm 0.03$ &  MDM 1.3 m\\
2002 Oct 5.195 & $B$ & $20.12 \pm 0.03$ &  MDM 1.3 m\\ 
2002 Oct 5.211 & $B$ & $20.17 \pm 0.04$ &  MDM 1.3 m\\
2002 Oct 5.227 & $B$ & $20.21 \pm 0.04$ &  MDM 1.3 m\\ 
2002 Oct 5.248 & $B$ & $20.22 \pm 0.03$ &  MDM 1.3 m\\ 
2002 Oct 5.265 & $B$ & $20.23 \pm 0.03$ &  MDM 1.3 m\\ 
2002 Oct 5.280 & $B$ & $20.18 \pm 0.04$ &  MDM 1.3 m\\ 
2002 Oct 5.297 & $B$ & $20.32 \pm 0.04$ &  MDM 1.3 m\\ 
2002 Oct 5.313 & $B$ & $20.22 \pm 0.03$ &  MDM 1.3 m\\ 
2002 Oct 5.329 & $B$ & $20.27 \pm 0.03$ &  MDM 1.3 m\\
2002 Oct 5.345 & $B$ & $20.23 \pm 0.03$ &  MDM 1.3 m\\ 
2002 Oct 5.360 & $B$ & $20.15 \pm 0.03$ &  MDM 1.3 m\\
2002 Oct 5.376 & $B$ & $20.24 \pm 0.03$ &  MDM 1.3 m\\ 
2002 Oct 5.396 & $B$ & $20.28 \pm 0.03$ &  MDM 1.3 m\\ 
2002 Oct 5.411 & $B$ & $20.22 \pm 0.04$ &  MDM 1.3 m\\ 
2002 Oct 5.426 & $B$ & $20.25 \pm 0.04$ &  MDM 1.3 m\\ 
2002 Oct 5.453 & $B$ & $20.34 \pm 0.05$ &  MDM 1.3 m\\ 
2002 Oct 5.469 & $B$ & $20.27 \pm 0.05$ &  MDM 1.3 m\\ 
2002 Oct 5.485 & $B$ & $20.35 \pm 0.05$ &  MDM 1.3 m\\ 
2002 Oct 6.325 & $B$ & $21.03 \pm 0.02$ &  MDM 1.3 m\\ 
2002 Oct 7.318 & $B$ & $21.26 \pm 0.02$ &  MDM 1.3 m\\ 
2002 Oct 8.359 & $B$ & $21.66 \pm 0.03$ &  MDM 1.3 m\\ 
2002 Oct 8.484 & $B$ & $21.72 \pm 0.05$ &  MDM 1.3 m\\ 
2002 Oct 9.224 & $B$ & $21.90 \pm 0.03$ &  MDM 1.3 m\\ 
2002 Oct 11.303 & $B$ & $22.27 \pm 0.04$ &  MDM 1.3 m\\ 
2002 Oct 12.316 & $B$ & $22.52 \pm 0.11$ &  MDM 1.3 m\\ 
2002 Nov 27.19  & $B$ & $24.53 \pm 0.06$ & MDM 2.4 m\\ 
2002 Oct 5.123  & $V$ & $19.39 \pm 0.04$ &  MDM 1.3 m\\
2002 Oct 5.147 & $V$ & $19.42 \pm 0.07$ & MDM 1.3 m\\
2002 Oct 5.176 & $V$ & $19.52 \pm 0.02$ & MDM 1.3 m \\ 
2002 Oct 5.199 & $V$ & $19.53 \pm 0.03$ & MDM 1.3 m\\
2002 Oct 5.215 & $V$ & $19.56 \pm 0.03$ & MDM 1.3 m\\ 
2002 Oct 5.231 & $V$ & $19.57 \pm 0.03$ & MDM 1.3 m\\ 
2002 Oct 5.253 & $V$ & $19.57 \pm 0.03$ & MDM 1.3 m\\ 
2002 Oct 5.269 & $V$ & $19.62 \pm 0.03$ & MDM 1.3 m\\ 
2002 Oct 5.285 & $V$ & $19.61 \pm 0.03$ & MDM 1.3 m\\ 
2002 Oct 5.301 & $V$ & $19.69 \pm 0.03$ & MDM 1.3 m\\ 
2002 Oct 5.318 & $V$ & $19.60 \pm 0.03$ & MDM 1.3 m\\ 
2002 Oct 5.333 & $V$ & $19.62 \pm 0.03$ & MDM 1.3 m\\ 
2002 Oct 5.349 & $V$ & $19.66 \pm 0.03$ & MDM 1.3 m\\ 
2002 Oct 5.365 & $V$ & $19.59 \pm 0.03$ & MDM 1.3 m\\ 
2002 Oct 5.380 & $V$ & $19.58 \pm 0.03$ & MDM 1.3 m\\ 
2002 Oct 5.400 & $V$ & $19.66 \pm 0.03$ & MDM 1.3 m\\ 
2002 Oct 5.415 & $V$ & $19.75 \pm 0.03$ & MDM 1.3 m\\ 
2002 Oct 5.442 & $V$ & $19.67 \pm 0.03$ & MDM 1.3 m\\ 
2002 Oct 5.457 & $V$ & $19.70 \pm 0.03$ & MDM 1.3 m\\ 
2002 Oct 5.473 & $V$ & $19.73 \pm 0.04$ & MDM 1.3 m\\ 
2002 Oct 5.490 & $V$ & $19.73 \pm 0.04$ & MDM 1.3 m\\
2002 Oct 5.126 & $R$ & $18.91 \pm 0.03$  &  MDM 1.3 m\\
2002 Oct     5.150    & $R$ & $18.89 \pm    0.06$ &  MDM 1.3 m\\
 2002 Oct    5.185    & $R$ & $19.12 \pm   0.02$ &  MDM 1.3 m\\
 2002 Oct    5.202    & $R$ & $19.16 \pm  0.02$ &  MDM 1.3 m\\
 2002 Oct    5.218    & $R$ & $19.17 \pm   0.03$ &  MDM 1.3 m\\
  2002 Oct   5.235    & $R$ & $19.13 \pm   0.03$ &  MDM 1.3 m\\
  2002 Oct   5.257    & $R$ & $19.20 \pm   0.03$ &  MDM 1.3 m\\
  2002 Oct   5.274    & $R$ & $19.18 \pm   0.02$ &  MDM 1.3 m\\
   2002 Oct  5.289    & $R$ & $19.19 \pm   0.02$ &  MDM 1.3 m\\
   2002 Oct  5.305    & $R$ & $19.19 \pm   0.02$ &  MDM 1.3 m\\
   2002 Oct  5.321    & $R$ & $19.18 \pm   0.02$ &  MDM 1.3 m\\
  2002 Oct   5.337    & $R$ & $19.20 \pm   0.02$ &  MDM 1.3 m\\
   2002 Oct  5.353    & $R$ & $19.19 \pm   0.02$ &  MDM 1.3 m\\
   2002 Oct  5.368    & $R$ & $19.16 \pm   0.02$ &  MDM 1.3 m\\
  2002 Oct   5.384    & $R$ & $19.17 \pm   0.02$ &  MDM 1.3 m\\
  2002 Oct   5.403    & $R$ & $19.22 \pm   0.03$ &  MDM 1.3 m\\
  2002 Oct   5.419    & $R$ & $19.21 \pm   0.02$ &  MDM 1.3 m\\
  2002 Oct   5.445    & $R$ & $19.19 \pm   0.03$ &  MDM 1.3 m\\
  2002 Oct   5.461    & $R$ & $19.27 \pm  0.03$ &  MDM 1.3 m\\
  2002 Oct   5.476    & $R$ & $19.29 \pm   0.03$ &  MDM 1.3 m\\
  2002 Oct   5.493    & $R$ & $19.24 \pm   0.03$ &  MDM 1.3 m\\
  2002 Oct   6.112    & $R$ & $19.84 \pm   0.03$ &  MDM 1.3 m\\
  2002 Oct   6.294    & $R$ & $19.91 \pm   0.02$ &  MDM 1.3 m\\
  2002 Oct   6.485    & $R$ & $20.00 \pm   0.02$ &  MDM 1.3 m\\
  2002 Oct   7.110    & $R$ & $20.19 \pm   0.03$ &  MDM 1.3 m\\
   2002 Oct  7.276    & $R$ & $20.14 \pm   0.02$ &  MDM 1.3 m\\
   2002 Oct  7.472    & $R$ & $20.21 \pm   0.04$ &  MDM 1.3 m\\
  2002 Oct   8.295    & $R$ & $20.47 \pm   0.02$ &  MDM 1.3 m\\
  2002 Oct   8.427    & $R$ & $20.52 \pm   0.03$  &  MDM 2.4 m\\
  2002 Oct   9.182    & $R$ & $20.85 \pm   0.04$ &  MDM 1.3 m\\
  2002 Oct   9.334    & $R$ & $20.79 \pm   0.02$ &  MDM 2.4 m\\
  2002 Oct  10.298    & $R$ & $21.03 \pm   0.02$ &  MDM 2.4 m\\
  2002 Oct   11.258    & $R$ & $21.23 \pm   0.04$ &  MDM 1.3 m\\
  2002 Oct   11.401    & $R$ & $21.30 \pm   0.03$ &  MDM 2.4 m\\
  2002 Oct   12.267    & $R$ & $21.40 \pm   0.04$ &  MDM 1.3 m\\
  2002 Oct   12.330    & $R$ & $21.44 \pm   0.03$ &  MDM 2.4 m\\
  2002 Oct   15.297    & $R$ & $22.18 \pm   0.07$ &  MDM 2.4 m\\
  2002 Oct   16.330    & $R$ & $22.33 \pm   0.10$ &  MDM 2.4 m\\
  2002 Oct   25.270    & $R$ & $23.10 \pm   0.06$ &  MDM 2.4 m\\
  2002 Nov   25.125    & $R$ & $23.85 \pm  0.08$ &  MDM 2.4 m\\
  2002 Nov   26.177    & $R$ & $23.87 \pm   0.08$ &  MDM 2.4 m\\
2002 Oct   5.130 & $I$ & $18.42 \pm 0.07$  & MDM 1.3 m\\
2002 Oct 5.155  & $I$ & $18.40 \pm 0.08$ & MDM 1.3 m\\
2002 Oct 5.191  & $I$ & $18.46 \pm 0.03$ & MDM 1.3 m\\
2002 Oct 5.208  & $I$ & $18.45 \pm 0.03$ & MDM 1.3 m\\
2002 Oct 5.223  & $I$ & $18.55 \pm 0.05$ & MDM 1.3 m\\
2002 Oct 5.240  & $I$ & $18.55 \pm 0.04$ & MDM 1.3 m\\
2002 Oct 5.261  & $I$ & $18.54 \pm 0.04$ & MDM 1.3 m\\
2002 Oct 5.278  & $I$ & $18.54 \pm 0.03$ & MDM 1.3 m\\
2002 Oct 5.293  & $I$ & $18.56 \pm 0.04$ & MDM 1.3 m\\
2002 Oct 5.302  & $I$ & $18.57 \pm 0.04$ & MDM 1.3 m\\
2002 Oct 5.325  & $I$ & $18.57 \pm 0.04$ & MDM 1.3 m\\
2002 Oct 5.341  & $I$ & $18.53 \pm 0.04$ & MDM 1.3 m\\
2002 Oct 5.357  & $I$ & $18.55 \pm 0.03$ & MDM 1.3 m\\
2002 Oct 5.372  & $I$ & $18.53 \pm 0.04$ & MDM 1.3 m\\
2002 Oct 5.392  & $I$ & $18.54 \pm 0.03$ & MDM 1.3 m\\
2002 Oct 5.408  & $I$ & $18.59 \pm 0.04$ & MDM 1.3 m\\
2002 Oct 5.423  & $I$ & $18.54 \pm 0.05$ & MDM 1.3 m\\
2002 Oct 5.449  & $I$ & $18.67 \pm 0.04$ & MDM 1.3 m\\
2002 Oct 5.465  & $I$ & $18.68 \pm 0.05$ & MDM 1.3 m\\
2002 Oct 5.481  & $I$ & $18.65 \pm 0.05$ & MDM 1.3 m\\
2002 Oct 5.497  & $I$ & $18.61 \pm 0.05$ & MDM 1.3 m\\
\enddata
\end{deluxetable}
\clearpage

\begin{deluxetable}{ccccc}
\tablenum{2}
\tablewidth{0pt}
\tablecaption{LRIS Line Identifications}
\tablehead
{
Line($\lambda_{\rm vac}$(\AA)) & $\lambda_{\rm helio}$(\AA) & $z$ &  $f_{ij}$ &
$W_{\lambda}$(\AA)\tablenotemark{a}}
\startdata
Ly $\delta$(949.74 \AA)  & 3261.08  & 2.328 & 1.39$\times 10^{-2}$ & ... \\
Ly $\gamma$(972.54 \AA)  & 3203.94  & 2.294 & 2.90$\times 10^{-2}$ & 2.04 $\pm$ 0.60 	 \\
C III(977.02 \AA) & 3214.88 & 2.290 &  7.62$\times 10^{-1}$ & 4.07 $\pm$ 0.84  \\
Ly $\gamma$(972.54 \AA)  & 3231.41  & 2.323 & 2.90$\times 10^{-2}$ & 3.11 $\pm$ 0.55 	  \\
C III(977.02 \AA) & 3247.57 & 2.324 &  7.62$\times 10^{-1}$ & 3.68 $\pm$ 0.62  \\
Ly $\beta$(1025.72 \AA)  & 3376.11  & 2.292 & 7.91$\times 10^{-2}$ & 1.95 $\pm$ 0.55 	  \\
Ly $\beta$(1025.72 \AA)  & 3406.40 & 2.321 & 7.91$\times 10^{-2}$ & 7.17 $\pm$ 0.52	  \\
O VI(1031.93 \AA) & 3398.15 & 2.293 &  1.33$\times 10^{-1}$ & $\leq$ 1.02 	  \\
+O VI(1037.62 \AA)\tablenotemark{b} & 3416.88 & 2.293 & 6.61 $\times 10^{-2}$ & 	  \\
Si II(1194.75 \AA) & 3975.35 & 2.327 & 6.23$\times 10^{-1}$ & 2.37 $\pm$ 0.31	  \\
+Al II(1670.79 \AA)\tablenotemark{c} &         & 1.379 & 1.83 & 3.37 $\pm$ 0.43	  \\
--  & 3613.91 & -- & -- & --	  \\
--  & 3626.13 & -- & -- & --	  \\
--  & 3667.12 & -- & -- & --	  \\
--  & 3680.84 & -- & -- & --	  \\
Ly $\alpha$(1215.67 \AA)  & 4006.11 & 2.295 & 4.16$\times 10^{-1}$ & 3.91 $\pm$ 0.57	  \\
Ly $\alpha$(1215.67 \AA)  & 4006.11 & 2.295 & 4.16$\times 10^{-1}$ & 3.91 $\pm$ 0.57	  \\
Ly $\alpha$(1215.67 \AA)  & 4034.87 & 2.319 & 4.16$\times 10^{-1}$ & 4.82 $\pm$ 0.60	  \\
Ly $\alpha$(1215.67 \AA)  & 4046.24 & 2.328 & 4.16$\times 10^{-1}$ & emission line \\
N V (1238.82 \AA) & 4079.37 & 2.293 & 1.57$\times 10^{-1}$ & $\leq$ 0.37	  \\
+N V (1242.80 \AA)\tablenotemark{b} & 4092.54        & 2.293 &  7.82$\times 10^{-2}$ & 	  \\
Al II(1670.79 \AA) & 4345.80 & 1.601 & 1.83 & 0.80 $\pm$ 0.24	  \\
Si IV(1393.76 \AA) & 4590.26 & 2.293 & 5.14$\times 10^{-1}$ & 0.46 $\pm$ 0.10	 \\
Si IV(1393.76 \AA) & 4623.72 & 2.317 & 5.14$\times 10^{-1}$ & 1.33 $\pm$ 0.30	 \\
+Si IV(1402.77 \AA)\tablenotemark{b} &         & 2.296 & 2.55$\times 10^{-1}$ &  \\
Si IV(1393.76 \AA) & 4632.06 & 2.323 & 5.14$\times 10^{-1}$ & 1.14 $\pm$ 0.47	 \\
Si IV(1402.77 \AA) & 4653.64 & 2.317 & 2.55$\times 10^{-1}$ & 0.81 $\pm$ 0.15	 \\
Si IV(1402.77 \AA) & 4662.02 & 2.323 & 2.55$\times 10^{-1}$ & 1.01 $\pm$ 0.33	 \\
C IV(1548.20 \AA) & 5096.29 & 2.292 & 1.91$\times 10^{-1}$  & 0.96 $\pm$ 0.22    \\
C IV(1550.77 \AA) & 5105.29 & 2.292 & 9.52$\times 10^{-2}$  & 0.75 $\pm$ 0.16    \\
\enddata
\end{deluxetable}
\clearpage

\begin{deluxetable}{ccccc}
\tablenum{2}
\tablewidth{0pt}
\tablecaption{LRIS Line Identifications (Continued)  }
\tablehead
{
Line($\lambda_{\rm vac}$(\AA)) & $\lambda_{\rm helio}$(\AA) & $z$ & $f_{ij}$ &
$W_{\lambda}$(\AA)\tablenotemark{a}}
\startdata
C IV(1548.20 \AA) & 5134.77 & 2.317 & 1.91$\times 10^{-1}$  & 1.71 $\pm$ 0.46    \\
C IV(1548.20 \AA) & 5143.23 & 2.322 & 1.91$\times 10^{-1}$  & 2.02 $\pm$ 0.51    \\
+C IV(1550.77 \AA)\tablenotemark{b} &  & 2.317 & 9.52$\times 10^{-2}$  &     \\
C IV(1550.77 \AA) & 5152.37 & 2.322 & 9.52$\times 10^{-2}$  & 1.71 $\pm$ 0.45    \\
Al II(1670.79 \AA) & 5559.70 & 2.328 & 1.83 & 0.72 $\pm$ 0.16	  \\
Fe II(2374.46 \AA) & 5652.60 & 1.381 & 3.26$\times 10^{-2}$ & 0.64 $\pm$ 0.23	 \\
Fe II(2344.21 \AA) & 6101.00 & 1.603 & 1.10$\times 10^{-1}$ & 0.56 $\pm$ 0.17	 \\
Fe II(2586.65 \AA) & 6156.23 & 1.380 & 6.84$\times 10^{-2}$ & 0.82 $\pm$ 0.21	 \\
Fe II(2374.46 \AA) & 6175.49 & 1.601 & 3.26$\times 10^{-2}$ & 0.99 $\pm$ 0.17	 \\
+ Al III(1854.72 \AA)\tablenotemark{c} &      & 2.330 & 5.60$\times 10^{-1}$ & 0.77 $\pm$ 0.13  \\
Fe II(2600.17 \AA) & 6188.73 & 1.380 & 2.24$\times 10^{-1}$ & 0.94 $\pm$ 0.29	 \\
Fe II(2382.77 \AA) & 6201.15 & 1.602 & 3.01$\times 10^{-1}$ & 1.12 $\pm$ 0.32	  \\
+ Al III(1862.79 \AA)\tablenotemark{c} &      & 2.329 & 2.79$\times 10^{-1}$ & 0.88 $\pm$ 0.25     \\
Mg II(2796.35 \AA) & 6656.10 & 1.380 & 6.12$\times 10^{-1}$& 1.81 $\pm$ 0.37	  \\
Mg II(2803.53 \AA) & 6672.88 & 1.380 & 3.05$\times 10^{-1}$ & 1.47 $\pm$ 0.32	  \\
Fe II(2586.65 \AA) & 6731.85 & 1.603 & 6.84$\times 10^{-2}$ & 0.68 $\pm$ 0.16	 \\
Fe II(2600.17 \AA) & 6766.45 & 1.602 & 2.24$\times 10^{-1}$ & 0.83 $\pm$ 0.26	 \\
Mg II(2796.35 \AA) & 7276.72 & 1.602 & 6.12$\times 10^{-1}$& 1.53 $\pm$ 0.37	  \\
Mg II(2803.53 \AA) & 7295.44 & 1.602 & 3.05$\times 10^{-1}$ & 1.30 $\pm$ 0.32	  \\
Mg I(2852.96 \AA) & 7423.74 & 1.602 & 1.83 &  0.45 $\pm$ 0.13  \\
Fe II(2344.21 \AA) & 7801.53 & 2.328 & 1.10$\times 10^{-1}$ & $\leq$ 0.39	 \\
Fe II(2382.77 \AA) & 7929.86 & 2.328 & 3.01$\times 10^{-1}$ & $\leq$ 0.59 	  \\
Fe II(2600.17 \AA) & 8653.37 & 2.328 & 2.24$\times 10^{-1}$ & 	$\leq$ 0.88  \\

\enddata
\tablenotetext{a}{Equivalent width in the rest frame.}
\tablenotetext{b}{Doublet or blended lines.}
\tablenotetext{c}{Alternative identification to previous entry.}
\end{deluxetable}
\clearpage

\clearpage

\begin{deluxetable}{ccccc}
\tablenum{3}
\tablewidth{0pt}
\tablecaption{Derived Column Densities for the $z_{3}$ System}
\tablehead
{
Ion($j$)  &  $\lambda_{\rm vac}$(\AA) & $f_{ij}$ &
log$(N_{j}$) & Component}
\startdata
Ly $\gamma$ & 972.54 & 2.90$\times 10^{-2}$ & 16.11  $\pm$ 0.07 & $z_{3A,B}$\\
            &  &  & 15.92  $\pm$ 0.12 & $z_{3C}$\\
Ly $\beta$ & 1025.72 & 7.91$\times 10^{-2}$ & 15.99  $\pm$  0.03 & $z_{3A,B}$\\
 &  &  & 15.42  $\pm$  0.11 & $z_{3C}$\\
Ly $\alpha$ & 1215.67 & 4.16$\times 10^{-1}$ & 14.95  $\pm$  0.05 
& $z_{3A,B}$\\
 &  &  & 14.86  $\pm$ 0.06 &   $z_{3C}$\\
C III & 977.02  & 7.62$\times 10^{-1}$  & 14.76 $\pm$ 0.06 &  $z_{3A,B}$  \\
      &   &     & 	14.80 $\pm$ 0.08 &  $z_{3C}$  \\
C IV & 1548.20 & 1.91$\times 10^{-1}$ & 14.46 $\pm$ 0.15 &   $z_{3A}$\\
     &         &       & 14.63 $\pm$ 0.09  &   $z_{3B}$\\
     &         &       & 14.38 $\pm$ 0.08 &   $z_{3C}$\\
C IV & 1550.77 & 9.52$\times 10^{-2}$ & 14.93 $\pm$ 0.10 & $z_{3A}$\\
     &         &  & 14.63 $\pm$ 0.21 & $z_{3B}$\\
     &         &  & 14.57 $\pm$ 0.08  & $z_{3C}$\\
N V & 1238.82 & 1.57$\times 10^{-1}$ & $\leq$ 14.23  &  $z_{3C}$\\
N V & 1242.80 & 7.82$\times 10^{-2}$ & $\leq$ 14.54  &  $z_{3C}$\\
O VI & 1031.93  & 1.33$\times 10^{-1}$ & $\leq$ 14.91 	&  $z_{3C}$\\
O VI & 1037.62  & 6.61 $\times 10^{-2}$ & $\leq$ 15.21 & $z_{3C}$\\
Si IV & 1393.76 & 5.14$\times 10^{-1}$ & 14.11 $\pm$ 0.15 &  $z_{3A}$\\
      &         &       & 14.10 $\pm$ 0.10 &   $z_{3B}$\\
      &         &       & 13.72 $\pm$ 0.08 & $z_{3C}$\\
Si IV & 1402.77 & 2.55$\times 10^{-1}$ & 14.36 $\pm$  0.12 &   $z_{3A}$\\
      &         &       & 14.26 $\pm$  0.07 & $z_{3B}$\\
      &         &       & 13.71 $\pm$ 0.16  & $z_{3C}$\\
\enddata
\end{deluxetable}

\clearpage

\begin{deluxetable}{ccc}
\tablenum{4}
\tablewidth{0pt}
\tablecaption{Total Ionic Abundances in the $z_{3}$ System}
\tablehead
{
Ion($j$)  &
log$(N_{j})$ & Component}
\startdata
H$^{0}$ & 16.11  $\pm$ 0.31 & $z_{3A,B}$\\
  & 15.92  $\pm$ 0.26 & $z_{3C}$\\
C$^{+3}$ & $\geq$ 15.05 &  $z_{3A}$  \\
  & $\geq$ 14.93  &   $z_{3B}$\\
  & 15.09 $\pm$ 0.08 &   $z_{3C}$\\
N$^{+4}$  & $\leq$ 14.71  &   $z_{3C}$\\
O$^{+5}$  & $\leq$ 15.39  &   $z_{3C}$\\
Si$^{+3}$ & 14.55 $\pm$ 0.13 & $z_{3A}$\\
   & 14.49 $\pm$ 0.08 &   $z_{3B}$\\
   & 14.02 $\pm$ 0.12 & $z_{3C}$\\
\enddata
\end{deluxetable}
\clearpage

\begin{figure}[tbp] \label{lris} \figurenum{1}
\begin{center}
\epsscale{.90}
\plotone{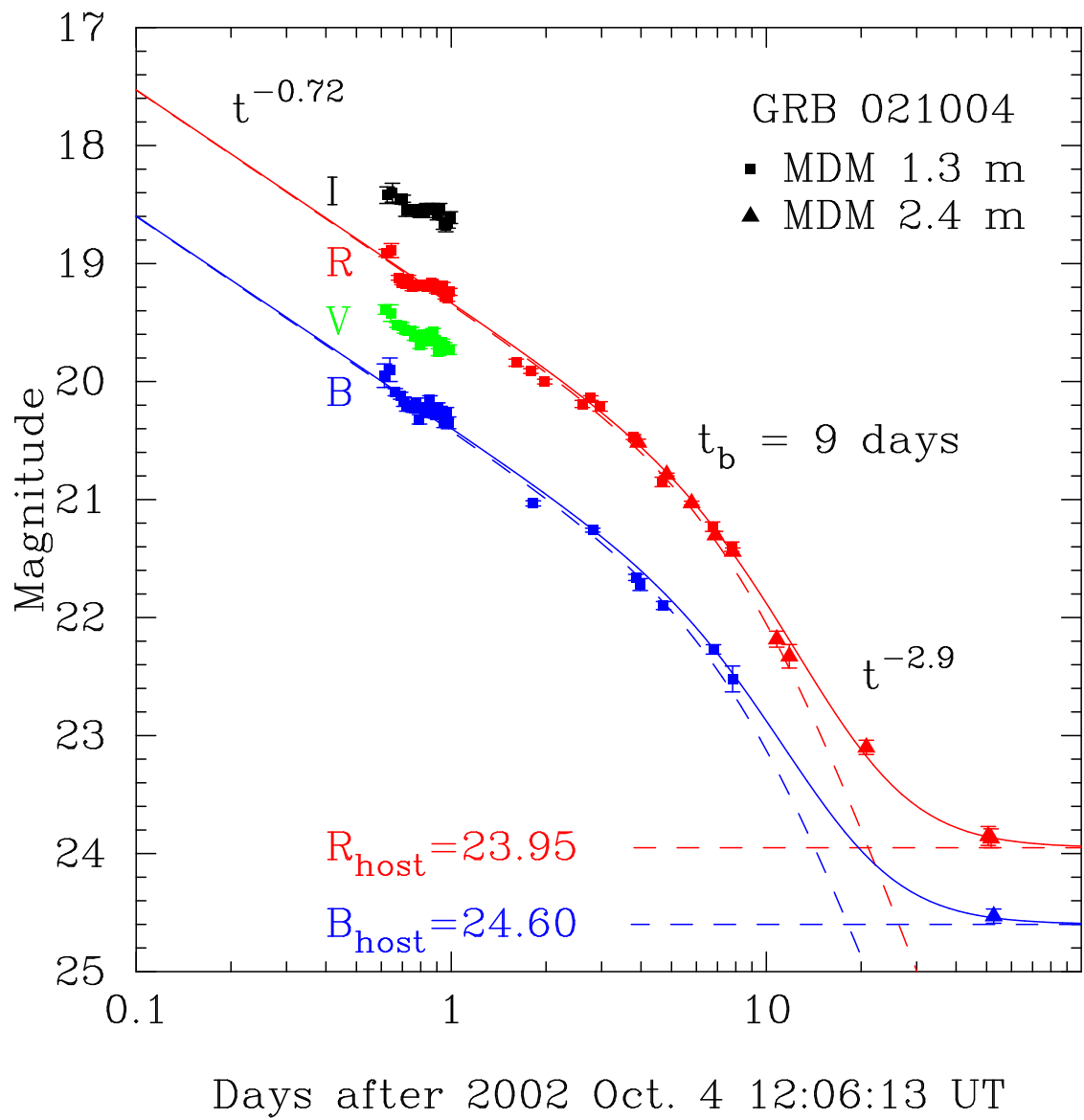}
\caption{
Monitoring of GRB 021004 from the MDM Observatory.
A broken power-law fit to the decay, including a constant 
contribution from a blue host galaxy, is shown. 
The data are listed in Table 1.
}
\end{center}
\end{figure}
\clearpage

\begin{figure}[tbp] \label{lris} \figurenum{2}
\begin{center}
\epsscale{.90}
\plotone{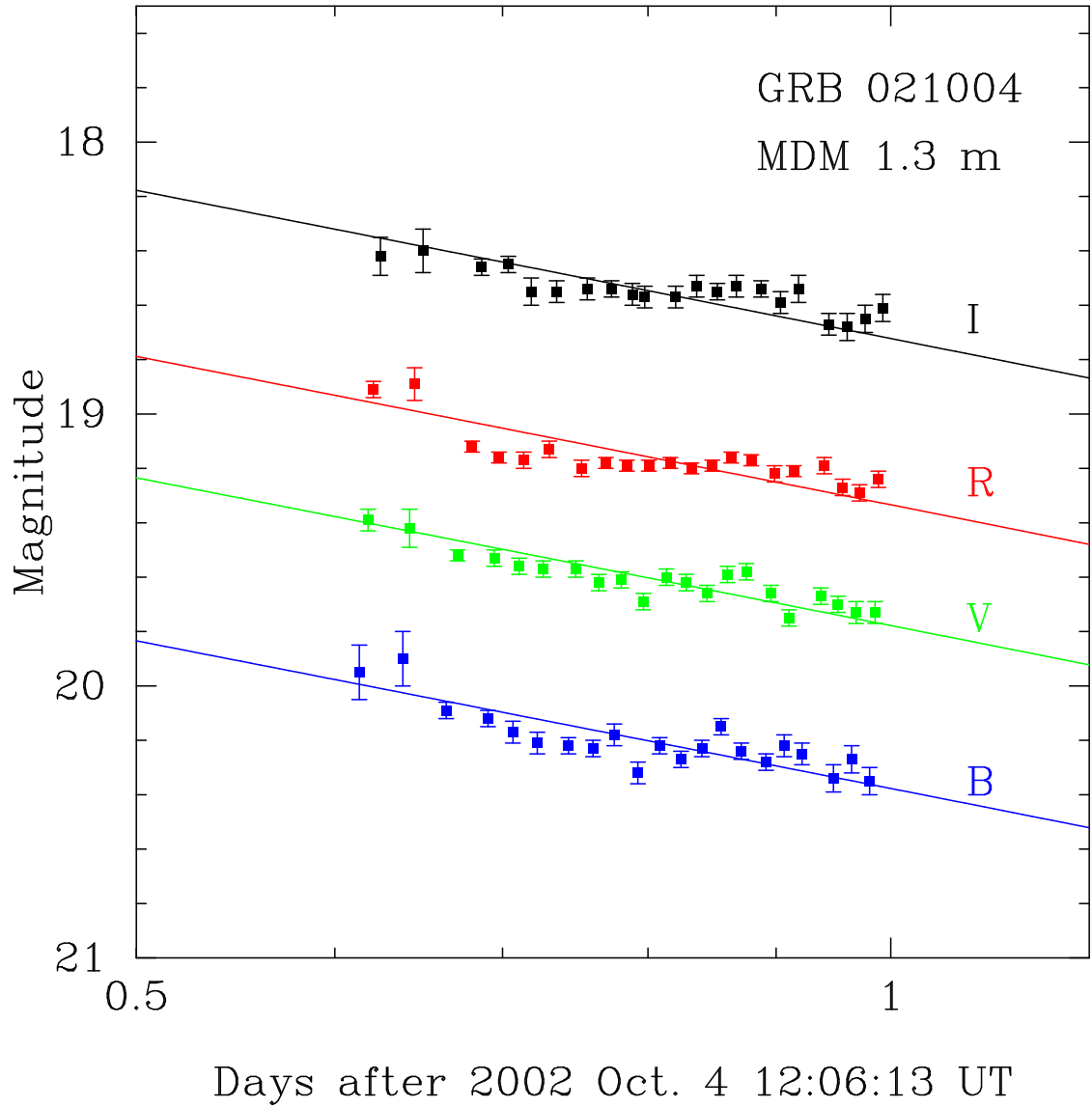}
\caption{
Multicolor light curves of GRB 021004 obtained during the first
day show significant deviations from a mean power-law
decay.
}
\end{center}
\end{figure}
\clearpage

\begin{figure}[tbp] \label{lris} \figurenum{3}
\begin{center}
\epsscale{.90}
\plotone{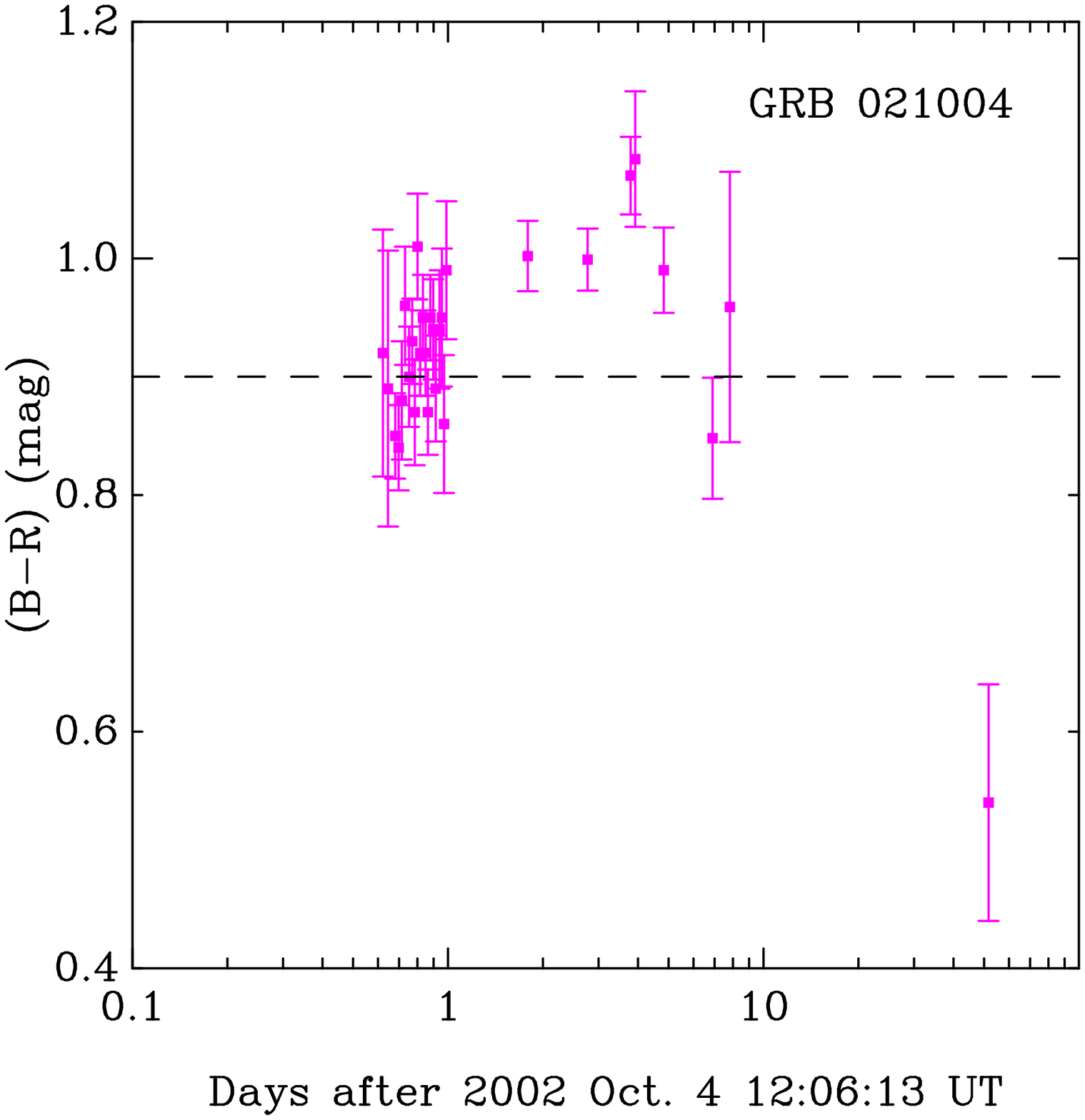}
\caption{
Color changes in the optical light curve of GRB 021004 represented as \br.
The data show the distinct color evolution of the afterglow. Late-time colors
are clearly contaminated by a blue host galaxy. 
}
\end{center}
\end{figure}
\clearpage

\begin{figure}[bp] \label{lris} \figurenum{4}
\begin{center}
\epsscale{.90}
\plotone{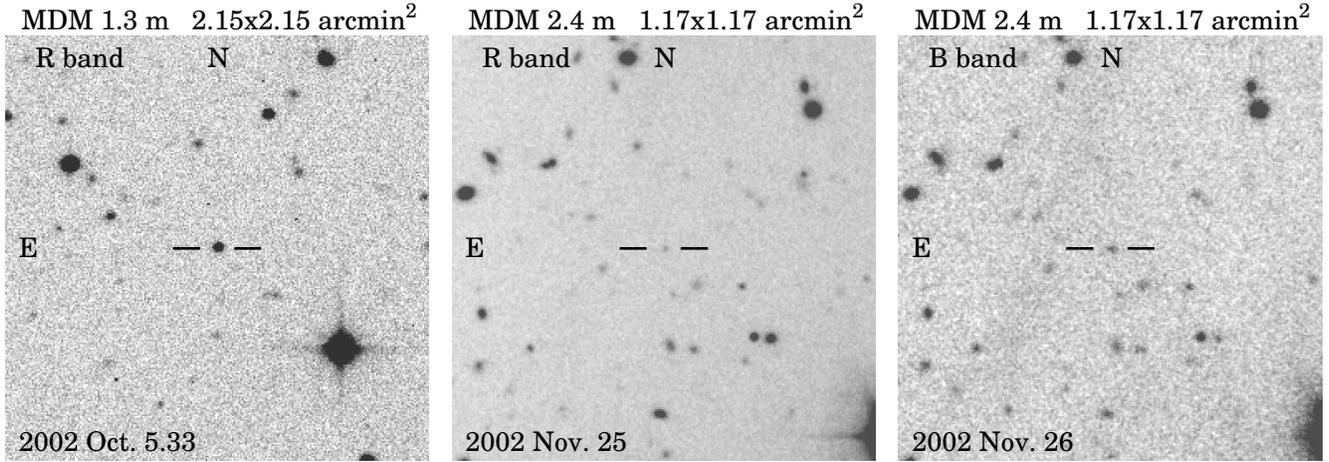}
\caption{
Early and late-time optical images
from the MDM Observatory. {\it Left:} $R = 19.32 \pm 0.02$ mag at 
$t = 19.8$ hours.
{\it Middle:} $R = 23.95 \pm 0.08$ mag at $t \approx 51$ days.
{\it Right:} $B = 24.60 \pm 0.06$ mag at $t \approx 52$ days.  The host galaxy
therefore has $B-R \approx 0.65$ mag, which is bluer than the 
afterglow ($B-R \approx 1.05$ mag), and bluer than the surrounding galaxies.
}
\end{center}
\end{figure}
\clearpage

\begin{figure}[bp] \label{lris} \figurenum{5}
\begin{center}
\epsscale{.90}
\plotone{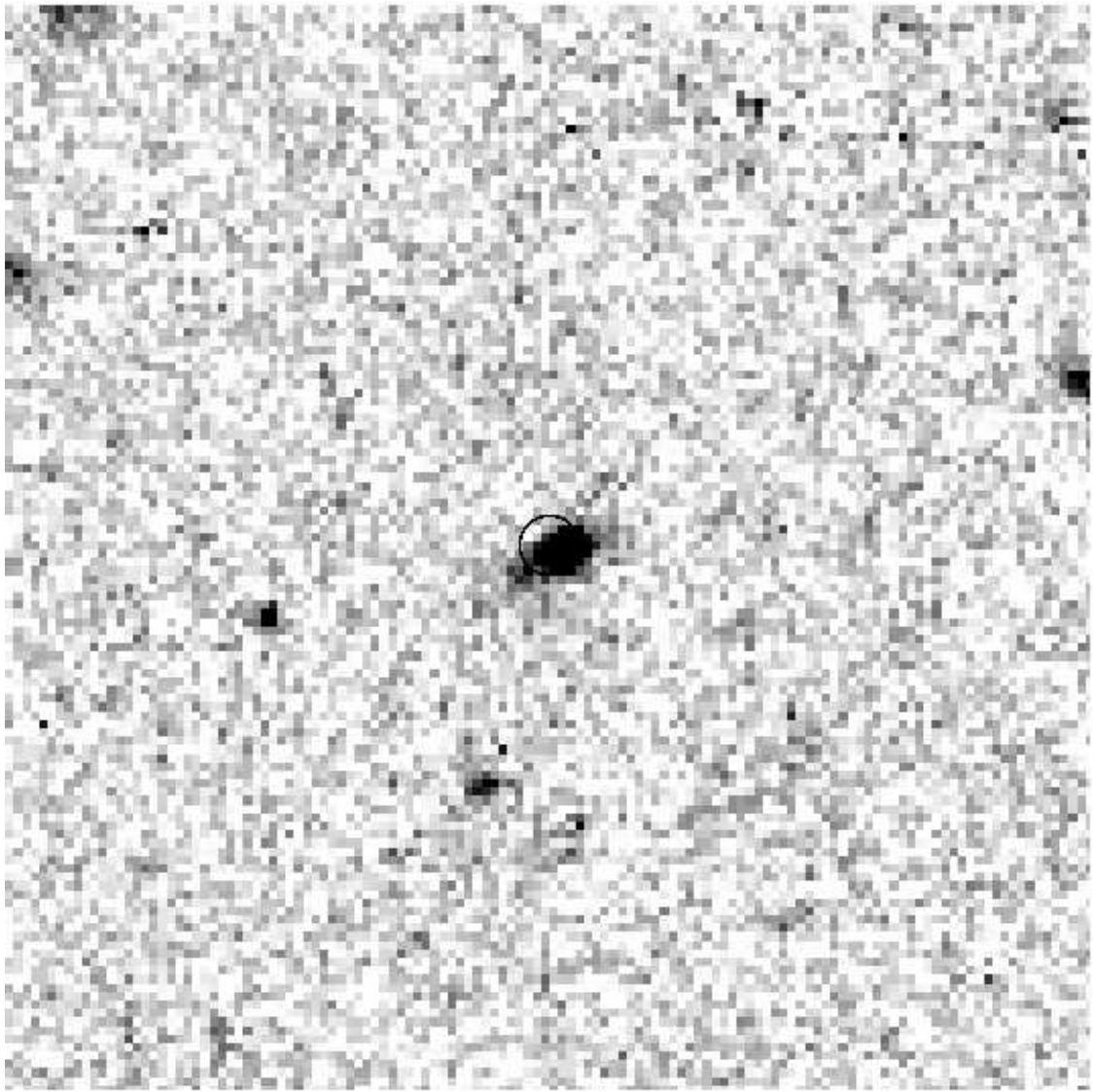}
\caption{
{\it HST\/} ACS F606W image of the GRB 021004 host galaxy on
2002 Nov. 26. The astrometric position of the 
OT ({\it circle}) was determined from an earlier ACS epoch obtained on
2002 Oct. 11 when the OT
dominated the light. North is up, and east is to the left. 
The field is $6^{\prime\prime}$ across and the error circle is drawn
with a $0^{\prime\prime}\!.15$ radius.}
\end{center}
\end{figure}
\clearpage

\begin{figure}[tbp] \label{lris} \figurenum{6}
\begin{center}
\epsscale{.90}
\plotone{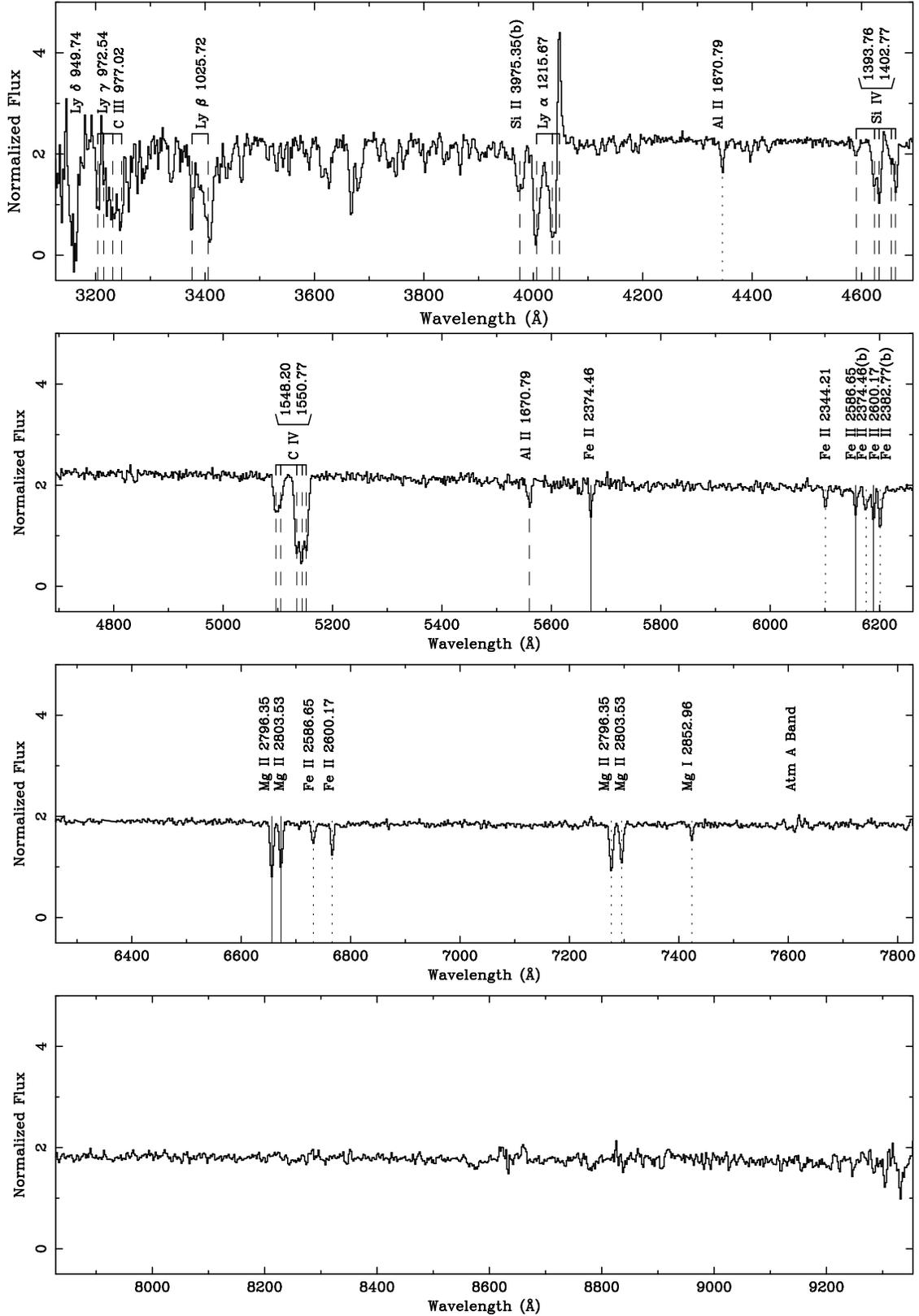}
\caption{
LRIS spectrum of the GRB 021004 afterglow on 2002 Oct. 8.507. Three
absorption systems are labeled $z_{1}=1.380$ (solid lines), 
$z_{2}=1.602$ (dotted lines), and $z_{3}=2.328$ (dashed lines).
}
\end{center}
\end{figure}
\clearpage

\begin{figure}[tbp] \label{lris} \figurenum{7}
\begin{center}
\epsscale{.80}
\plotone{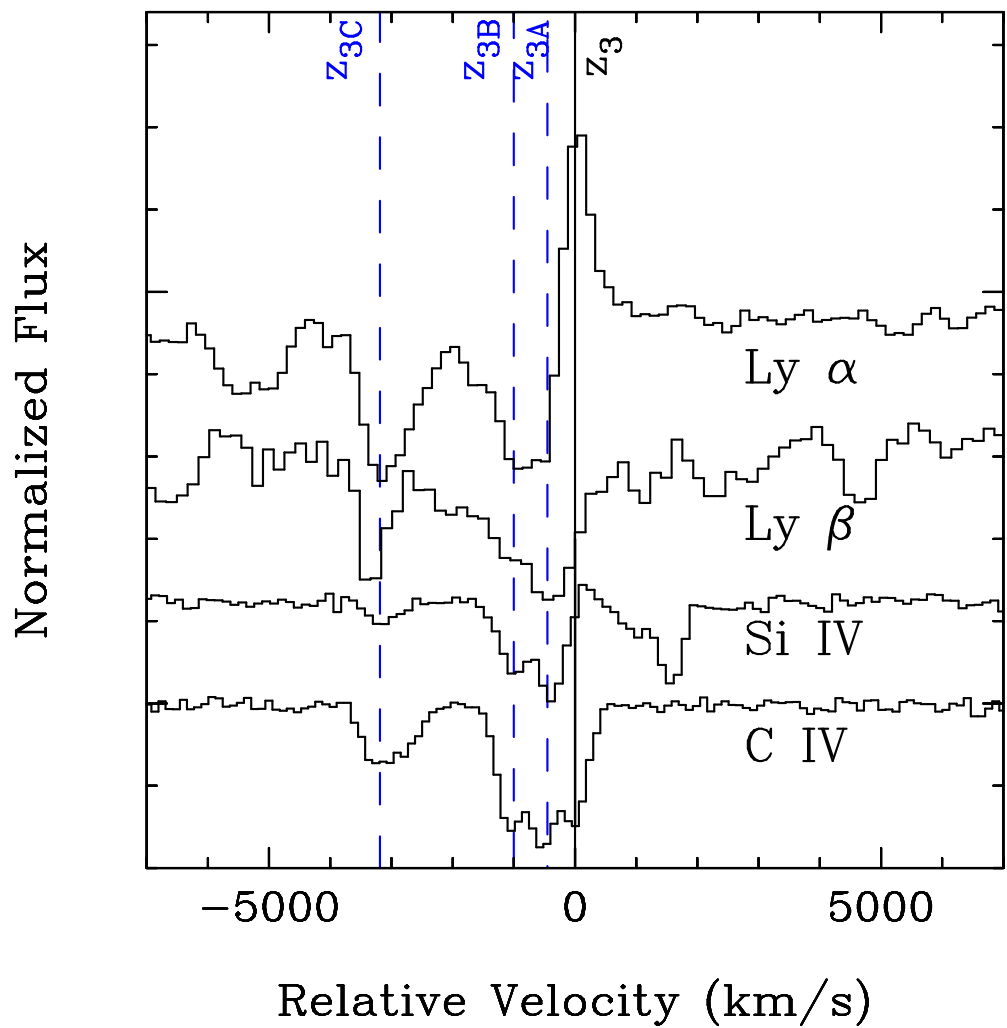}
\caption{
Blueshifted Lyman-$\alpha$, Lyman-$\beta$, Si IV, and C IV 
 absorbers in the GRB 021004 
afterglow spectrum plotted 
in velocity space.  As zero velocity we use the systemic redshift 
$z_{3}=2.328$. 
The dashed lines indicate blueshifted absorbers
 at $z_{3A}=2.323$, $z_{3B}=2.317$, and $z_{3C}=2.293$.
}
\end{center}
\end{figure}
\clearpage

\begin{figure}[tbp] \label{lris} \figurenum{8}
\begin{center}
\epsscale{1.00}
\plotone{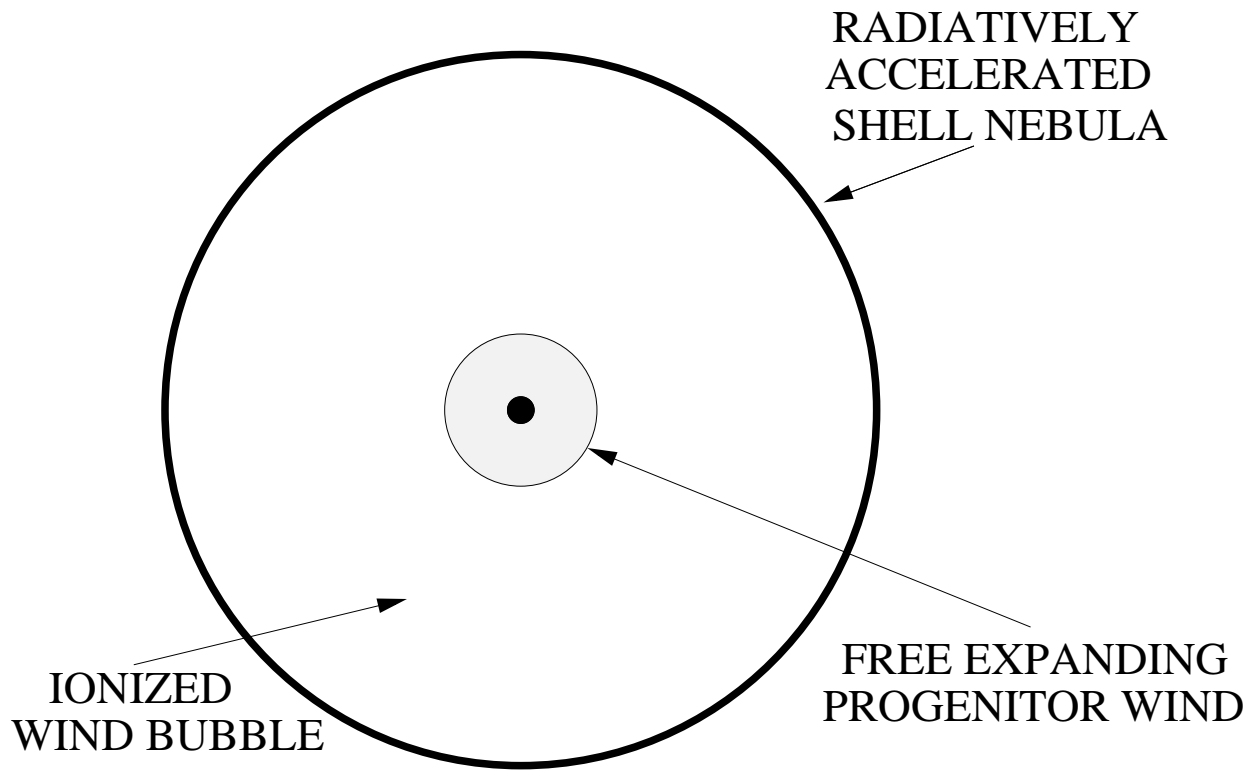}
\caption{
Schematic cross-section of a shell-nebula structure with 
various features including the termination of the wind
and the central star. The model cannot reproduce the great wealth of 
structure observed within shell nebulae.} 
\end{center}
\end{figure}
\clearpage

\begin{figure}[tbp] \label{lris} \figurenum{9}
\begin{center}
\epsscale{1.00}
\plotone{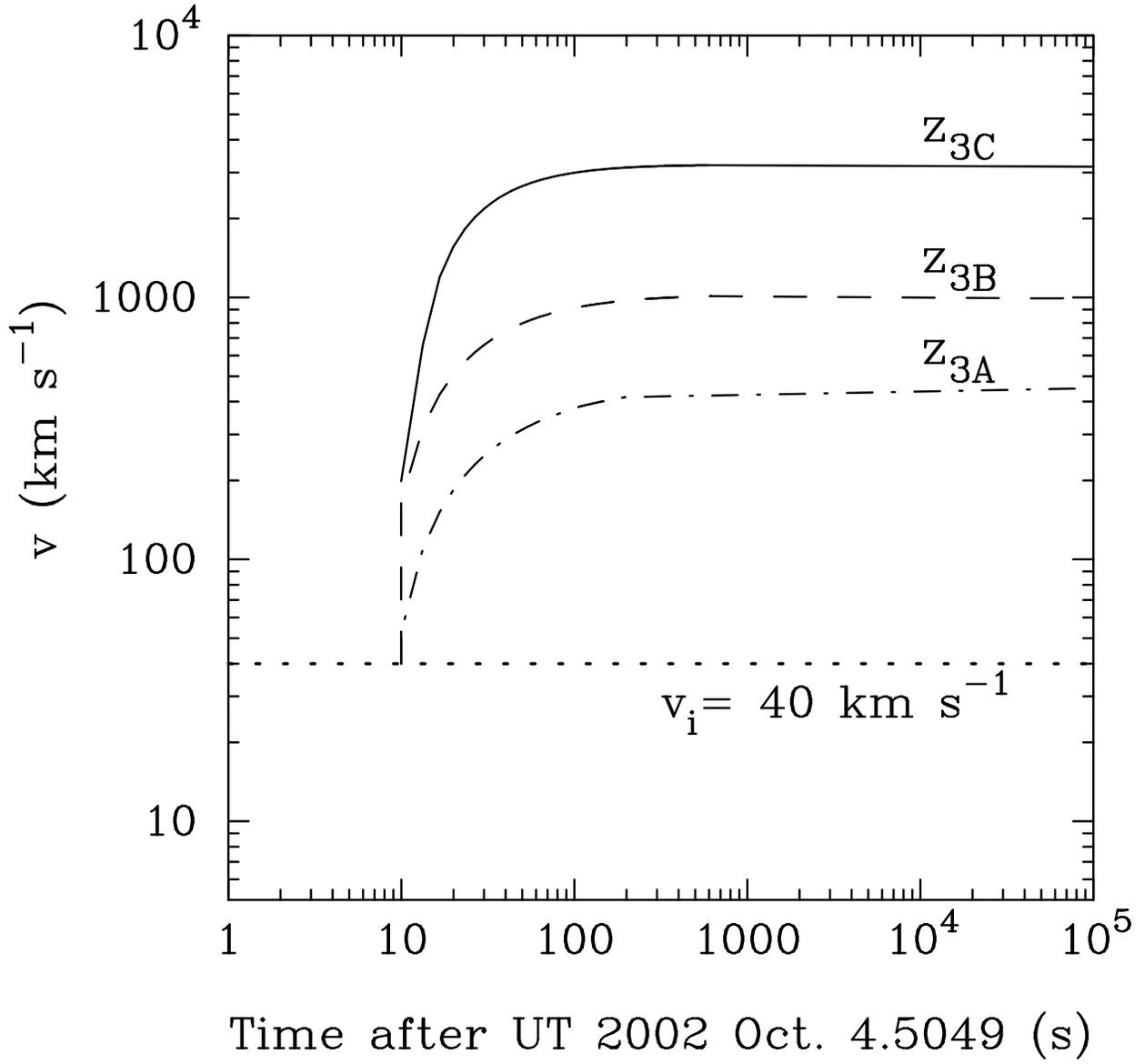}
\caption{
Total velocity as a function of time 
acquired by shell-nebula fragments located at
 0.3 pc, 0.54 pc, and 0.8 pc for $z_{3A}$, $z_{3B}$, and $z_{3C}$ 
respectively. The dotted
line corresponds to an initial velocity of expansion of the shell
nebula $v_{i} \approx 40$ km~s$^{-1}$. 
}
\end{center}
\end{figure}
\clearpage

\end{document}